\documentclass[acmsmall, nonacm=true, printfolios=true]{acmart}
\pdfoutput=1

\definecolor{rwthblue}   {RGB}{  0,  84, 159}
\definecolor{rwthblue75}{RGB}{ 64, 127, 183}
\definecolor{rwthblue50}{RGB}{142, 186, 229}
\definecolor{rwthblue25}{RGB}{199, 221, 242}
\definecolor{rwthblue10}{RGB}{232, 241, 250}

\definecolor{rwthblack}  {RGB}{  0,   0,   0}
\definecolor{rwthblack75}{RGB}{100, 101, 103}
\definecolor{rwthblack50}{RGB}{156, 158, 159}
\definecolor{rwthblack25}{RGB}{207, 209, 210}
\definecolor{rwthblack10}{RGB}{236, 237, 237}

\definecolor{rwthdarkgray}{named}{rwthblack75}
\definecolor{rwthgray}{named}{rwthblack50}
\definecolor{rwthlightgray}{named}{rwthblack25}
\definecolor{rwthverylightgray}{named}{rwthblack10}

\definecolor{rwthmagenta}  {RGB}{227,   0, 102}
\definecolor{rwthmagenta75}{RGB}{233,  96, 136}
\definecolor{rwthmagenta50}{RGB}{241, 158, 177}
\definecolor{rwthmagenta25}{RGB}{249, 210, 218}
\definecolor{rwthmagenta10}{RGB}{253, 238, 240}

\definecolor{rwthyellow}  {RGB}{255, 237,   0}
\definecolor{rwthyellow75}{RGB}{255, 240,  85}
\definecolor{rwthyellow50}{RGB}{255, 245, 155}
\definecolor{rwthyellow25}{RGB}{255, 250, 209}
\definecolor{rwthyellow10}{RGB}{255, 253, 238}

\definecolor{rwthpetrol}   {RGB}{  0,  97, 101}
\definecolor{rwthpetrol75}{RGB}{ 45, 127, 131}
\definecolor{rwthpetrol50}{RGB}{125, 164, 167}
\definecolor{rwthpetrol25}{RGB}{191, 208, 209}
\definecolor{rwthpetrol10}{RGB}{230, 236, 236}

\definecolor{rwthturquoie}   {RGB}{  0, 152, 161}
\definecolor{rwthturquoise75}{RGB}{  0, 177, 183}
\definecolor{rwthturquoise50}{RGB}{137, 204, 207}
\definecolor{rwthturquoise25}{RGB}{202, 231, 231}
\definecolor{rwthturquoise10}{RGB}{235, 246, 246}

\definecolor{rwthgreen}  {RGB}{ 87, 171,  39}
\definecolor{rwthgreen75}{RGB}{141, 192,  96}
\definecolor{rwthgreen50}{RGB}{184, 214, 152}
\definecolor{rwthgreen25}{RGB}{221, 235, 206}
\definecolor{rwthgreen10}{RGB}{242, 247, 236}

\definecolor{rwthlightgreen}   {RGB}{189, 205,   0}
\definecolor{rwthlightgreen75}{RGB}{208, 217,  92}
\definecolor{rwthlightgreen50}{RGB}{224, 230, 154}
\definecolor{rwthlightgreen25}{RGB}{240, 243, 208}
\definecolor{rwthlightgreen10}{RGB}{249, 250, 237}

\definecolor{rwthorange}  {RGB}{246, 168,   0}
\definecolor{rwthorange75}{RGB}{250, 190,  80}
\definecolor{rwthorange50}{RGB}{253, 212, 143}
\definecolor{rwthorange25}{RGB}{254, 234, 201}
\definecolor{rwthorange10}{RGB}{255, 247, 234}

\definecolor{rwthred}  {RGB}{204,   7,  30}
\definecolor{rwthred75}{RGB}{216,  92,  65}
\definecolor{rwthred50}{RGB}{230, 150, 121}
\definecolor{rwthred25}{RGB}{243, 205, 187}
\definecolor{rwthred10}{RGB}{250, 235, 227}

\definecolor{rwthbordeaured}   {RGB}{161,  16,  53}
\definecolor{rwthbordeauxred75}{RGB}{182,  82,  86}
\definecolor{rwthbordeauxred50}{RGB}{205, 139, 135}
\definecolor{rwthbordeauxred25}{RGB}{229, 197, 192}
\definecolor{rwthbordeauxred10}{RGB}{245, 232, 229}

\definecolor{rwthviolet}  {RGB}{ 97,  33,  88}
\definecolor{rwthviolet75}{RGB}{131,  78, 117}
\definecolor{rwthviolet50}{RGB}{168, 133, 158}
\definecolor{rwthviolet25}{RGB}{210, 192, 205}
\definecolor{rwthviolet10}{RGB}{237, 229, 234}

\definecolor{rwthpurple}  {RGB}{122, 111, 172}
\definecolor{rwthpurple75}{RGB}{155, 145, 193}
\definecolor{rwthpurple50}{RGB}{188, 181, 215}
\definecolor{rwthpurple25}{RGB}{222, 218, 235}
\definecolor{rwthpurple10}{RGB}{242, 240, 247}

\newcommand{\sym}[1]{\text{Symmetric}\left( #1 \right)}

\newcommand{\lotri}[1]{\text{lowerTriangular}\left( #1 \right)}
\newcommand{\uptri}[1]{\text{upperTriangular}\left( #1 \right)}

\newcommand{\diag}[1]{\text{Diagonal}\left( #1 \right)}

\usepackage{subcaption}

\usepackage{tikz}
\usetikzlibrary{positioning}
\usetikzlibrary{shapes.geometric}
\usetikzlibrary{shapes.multipart}
\usetikzlibrary{plotmarks}
\tikzset{tbox/.style={fill=rwthblack10, rounded corners}}
\tikzset{active/.style={fill=rwthblue50, rounded corners}}
\tikzset{barrow/.style={line width=0.2mm}}

\usepackage{pgfplots}
\usepgfplotslibrary{statistics}
\usepackage{pgfplotstable}
\pgfplotsset{compat=1.14}
\pgfplotsset{x tick label style={/pgf/number format/1000 sep=}}

\usepackage{listings}
\lstdefinestyle{mystyle}{
	language=Python,
	columns=[c]fullflexible,
	mathescape=true,
	tabsize=4,
	breaklines=true,
	lineskip=1pt,
	morekeywords={yield},
	escapeinside={(*}{*)},
	numbers=left,
}
\newcommand{\su}{{\footnotesize$\times$}}

\AtBeginDocument{%
  \providecommand\BibTeX{{%
    \normalfont B\kern-0.5em{\scshape i\kern-0.25em b}\kern-0.8em\TeX}}}

\setcopyright{acmcopyright}
\copyrightyear{2019}
\acmYear{2019}
\acmDOI{10.1145/1122445.1122456}

\acmJournal{TOMS}
\acmVolume{37}
\acmNumber{4}
\acmMonth{12}

\begin{document}

\title{Linnea: Automatic Generation of Efficient Linear Algebra Programs}

\author{Henrik Barthels}
\email{barthels@aices.rwth-aachen.de}

\author{Christos Psarras}
\email{psarras@aices.rwth-aachen.de}
\affiliation{%
  \institution{AICES, RWTH Aachen University}
  \streetaddress{Schinkelstr. 2}
  \city{Aachen}
  \postcode{52062}
  \country{Germany}
}

\author{Paolo Bientinesi}
\email{pauldj@cs.umu.se}
\affiliation{%
  \institution{Ume\aa{} Universitet}
  \streetaddress{Linnaeus v\"ag 49}
  \city{Ume\aa{}}
  \country{Sweden}
}


\begin{abstract}
The translation of linear algebra computations into efficient sequences of library calls is a non-trivial task that requires expertise in both linear algebra and high-performance computing. Almost all high-level languages and libraries for matrix computations (e.g., Matlab, Eigen) internally use optimized kernels such as those provided by BLAS and LAPACK; however, their translation algorithms are often too simplistic and thus lead to a suboptimal use of said kernels, resulting in significant performance losses.
In order to combine the productivity offered by high-level languages, and the performance of low-level kernels, we are developing Linnea, a code generator for linear algebra problems.
As input, Linnea takes a high-level description of a linear algebra problem; as output, it returns an efficient sequence of calls to high-performance kernels.
Linnea uses a custom best-first search algorithm to find a first solution in less than a second, and increasingly better solutions when given more time.
In 125 test problems, the code generated by Linnea almost always outperforms Matlab, Julia, Eigen and Armadillo, with speedups up to and exceeding 10\su{}.
\end{abstract}

\ccsdesc[500]{Computing methodologies~Linear algebra algorithms}
\ccsdesc[500]{Software and its engineering~Compilers}
\ccsdesc[500]{Software and its engineering~Domain specific languages}
\ccsdesc[300]{Mathematics of computing~Mathematical software}

\keywords{linear algebra, code generation}

\maketitle

\section{Introduction}

A common high-performance computing workflow to accelerate the execution of target application problems consists in first identifying a set of computational building blocks, and then engaging in extensive algorithmic and code optimization.  Although this process leads to sophisticated and highly-tuned code, the performance gains in the computational building blocks do not necessarily carry over to the high-level application problems that domain experts solve in their day-to-day work.

In the domain of linear algebra, significant effort is put into optimizing BLAS and LAPACK implementations for all the
different architectures and hardware generations, and  for operations such as matrix-matrix multiplication, nearly optimal efficiency rates are attained. However, we observe a decrease in the number of users that actually go through the tedious, error-prone and time consuming process of using directly said libraries by writing their code in C or Fortran; instead, languages and libraries such as Matlab, Julia, Eigen, and Armadillo, which offer a higher level of abstraction, are becoming more and more popular.
These languages and libraries allow users to input a linear algebra problem as an expression which closely resembles the
mathematical description; this expression is then internally mapped to lower level building blocks such as BLAS and LAPACK.
While human productivity is hence increased, it has been shown that this translation frequently results in suboptimal code \cite{psarras2019arxiv}.

The following examples illustrate some of the challenges that arise in the mapping from high-level expression to
low-level kernels. A straightforward translation of the assignment $y_k := H^\dag y + ( I_n - H^\dag H ) x_k$, which
appears in an image restoration application \cite{Tirer:2017uv}, will result in code containing one $\mathcal{O}(n^3)$
matrix-matrix multiplication to compute $H^\dag H$. In contrast, by means of distributivity, this assignment can be
rewritten as $y_k := H^\dag (y - H x_k) + x_k$, and computed with only $\mathcal{O}(n^2)$ matrix-vector
multiplications. The computation of the expression
$$\frac{k}{k-1}B_{k-1} (I_n - A^T W_k ((k-1)I_l + W_k^T A B_{k-1} A^T W_k)^{-1} W_k^T A B_{k-1} )\text{,}$$
which is part of a stochastic method for the solution of least squares problems \cite{Chung:2017ws}, can be sped up by
identifying that the subexpression $W_k^T A$ or its transpose $(A^T W_k)^T$ appear four times. Often times,
application experts possess domain knowledge that leads to better implementations. In $x := (A^T A + \alpha^2 I)^{-1}
A^T b$ \cite{Golub:2006hl}, since $\alpha > 0$, it can be inferred that $A^T A + \alpha^2 I$ is symmetric positive
definite; as a result, the linear system can always be solved by using the Cholesky factorization, which is less costly
than LU or LDL. Most languages and libraries either do not offer the means to specify such additional knowledge,
or do not automatically exploit it.

In this paper, we discuss Linnea, a prototype of a code generator
that automates the translation of the mathematical description of a linear algebra problem to an efficient sequence of calls to BLAS and LAPACK kernels.%
\footnote{Linnea is available at
\url{https://github.com/HPAC/linnea}.
}
Linnea is written in Python and targets mid-to-large scale linear algebra expressions, where problems are typically compute bound. 
It currently supports real-valued computations, and parallelism via multi-threaded kernels.
One of the advantages of Linnea is that all optimizations are performed symbolically, using rewrite rules and term replacement, so the generated programs are correct by construction.

As input, Linnea accepts a sequence of assignments, where the left-hand side is a single operand, and the right-hand side is an expression built from addition, multiplication, subtraction, transposition, and inversion.
As operands, matrices, vectors, and scalars can be used.
Operands can be annotated with the properties shown in Tab.~\ref{tab:properties}.
It is possible for operands to have more than one property, as long as they do not contradict one another. For instance, a matrix can be diagonal and SPD, which implies that all diagonal elements are positive.
An example of the input to Linnea is shown in Fig.~\ref{fig:inputexample}.
As building blocks, Linnea uses BLAS and LAPACK kernels, as well as a small number of code snippets for simple operations not supported by those libraries. As output, we decided to generate Julia code because it offers a good tradeoff between simplicity and performance: Low-level wrappers expose the full functionality of BLAS and LAPACK, while additional routines can be implemented easily without compromising performance \cite{Bezanson:2018ip}. Fig.~\ref{fig:outputexample} shows an example of the generated code for the assignment $x := W(A^T(AWA^T)^{-1}b-c)$, which comes from an optimization problem \cite{straszak2015}.

While Linnea was built having in mind users that are not experts in numerical linear algebra or in high-performance computing, it is nonetheless useful for experts too: It saves implementation time, and it serves as a starting point for the optimization of algorithms. Since Linnea generates code, it is---unlike other languages and libraries---transparent in the sense that users can verify how solutions are computed.
In addition, Linnea can also generate a description of how the input expression was rewritten to generate a specific algorithm, together with the costs of the individual kernels.

Experiments indicate that the code generated by Linnea usually outperforms Matlab, Julia, Eigen and Armadillo. At the same time, the code generation time is mostly in the order of a few seconds, that is, significantly faster than human experts.

\begin{table}
\caption{The properties supported by Linnea.}
\begin{tabular}{l@{\hspace{20pt}}l@{\hspace{20pt}}l}
\toprule
  diagonal & upper triangular & lower triangular \\
  symmetric & symmetric positive semi-definite & symmetric positive definite\\
  orthogonal & orthogonal rows & orthogonal columns\\
  permutation & unit diagonal (for triangular matrices) & positive (for scalars)\\
  full rank & non-singular & zero\\
  identity & &\\
\bottomrule
\end{tabular}
\label{tab:properties}
\end{table}

\begin{figure}
\lstset{
	basicstyle=\ttfamily,
}
\begin{minipage}{0.6\textwidth}
\hrule
\begin{lstlisting}
m = 1000
n = 5000

Matrix H(m, n) <FullRank>
Matrix Hd(n, m) <FullRank>
IdentityMatrix I_n(n, n)
ColumnVector y(m) <>
ColumnVector y_k(n) <>
ColumnVector x_k(n) <>

Hd = trans(H)*inv(H*trans(H))
y_k = Hd*y + (I_n - Hd*H)*x_k
\end{lstlisting}
\hrule
\end{minipage}
\caption{An example of the input to Linnea.}
\label{fig:inputexample}
\end{figure}

\begin{figure}
\lstset{
	basicstyle=\ttfamily,
	numbers=left,
	xleftmargin=2em,
	emph={end},
	emphstyle=\bfseries,
}
\begin{minipage}{0.6\textwidth} 
\hrule
\begin{lstlisting}
W = diag(W)
Acopy = Array{Float64}(undef, 1000, 2000)
blascopy!(1000*2000, A, 1, Acopy, 1)
for i = 1:size(A, 2);
    view(A, :, i)[:] .*= W[i];
end;
S = Array{Float64}(undef, 1000, 1000)
gemm!('N', 'T', 1.0, A, Acopy, 0.0, S)
potrf!('L', S)
trsv!('L', 'N', 'N', S, b)
trsv!('L', 'T', 'N', S, b)
gemv!('T', 1.0, Acopy, b, -1.0, c)
c .*= W
\end{lstlisting}
\hrule
\end{minipage}
\caption{The generated code for $x := W(A^T(AWA^T)^{-1}b-c)$. Variables were renamed for better readability. Lines 4--6 is one of the code snippets for operations not supported by BLAS and LAPACK; the multiplication of a full and a diagonal matrix.}
\label{fig:outputexample}
\end{figure}

\paragraph{Organization of the Article}
Sec.~\ref{sec:relatedwork} surveys the related work. The basic ideas behind the code generation in Linnea are introduced in Sec.~\ref{sec:synthesis}. Details of the implementation are discussed in Sec.~\ref{sec:implementation}. Experimental results are presented in Sec.~\ref{sec:experiments}. Sec.~\ref{sec:conclusion} concludes the paper.

\section{Related Work}
\label{sec:relatedwork}

\subsection{Code Generation}

The translation from the intermediate representation of a program in the form of an expression tree to machine
instructions is a problem closely related to that discussed in this paper. However, existing approaches using pattern matching and dynamic programming \cite{Aho:1976ga, aho1989}, as well as bottom-up rewrite systems (BURS) \cite{pelegri1988} solely focus on expressions containing basic operations directly supported by machine instructions. The two main objectives of code generation are to minimize the number of instructions and to use registers optimally. While there are approaches that generate optimal code for arithmetic expressions, considering associativity and commutativity \cite{sethi1970}, more complex properties of the underlying algebraic domain, for example distributivity, are usually not considered.

Instead of applying optimization passes sequentially, Equality Saturation (EQ) \cite{Tate:2009kz} is a compilation technique that constructs many equivalent programs simultaneously, stored in a single intermediate representation. Domain specific knowledge can be provided in the form of axioms. Equality Saturation is more general in its scope than Linnea, as it allows for control flow. So far, EQ has only been implemented for Java, and it is not clear how well it would scale with the large number of axioms required to encode the optimizations that Linnea carries out.

\subsection{Tools and Languages for Linear Algebra}

Presently, a range of tools are available for the computation of linear algebra expressions. At one end of the spectrum
there are the aforementioned high-level programming languages such as Matlab, Octave, Julia, R, and Mathematica. In
those languages, working code can be written within minutes, with little or no knowledge of numerical linear
algebra. However, the resulting code (possibly numerically unstable\footnote{Some systems compute the condition number for certain operations and give a warning if results may be inaccurate.}) usually achieves suboptimal
performance~\cite{psarras2019arxiv}. One of the reasons is that, with the exception of Julia, which supports matrix
properties in its type system, these tools rarely make it possible to express properties. A few Matlab functions exploit properties by inspecting matrix entries, a step which could be avoided
with a more general method to annotate operands with properties. Furthermore, if the inverse operator is used, an
explicit inversion takes place, even if the faster and numerically more stable solution would be to solve a linear
system instead~\cite[Sec. 13.1]{higham1996}; it is up to the user to rewrite the inverse in terms of operators, such as
the Matlab ``$\slash$'' and ``$\backslash$''~\cite{matlabdoc}, which solve linear systems.

At the other end of the spectrum there are C/Fortran libraries such as BLAS \cite{dongarra1990} and LAPACK
\cite{anderson1999}, which offer highly optimized kernels for basic linear algebra operations. However, the
translation of a target problem into an efficient sequence of kernel invocations is a lengthy and error-prone process that requires deep understanding of both numerical linear algebra and high-performance computing.

In between, there are expression template libraries such as Eigen \cite{eigenweb}, Blaze \cite{Iglberger:2012hb}, and
Armadillo \cite{sanderson2010}, which provide a domain-specific language integrated within C++. They offer a compromise
between ease of use and performance. The main idea is to improve performance by eliminating temporary operands. While
both high-level languages and libraries increase the accessibility, they almost entirely ignore domain knowledge, and
because of this, they frequently deliver slow code.

The Transfor program \cite{gomez1998} is likely the first translator of linear algebra problems (written in Maple) into BLAS kernels; since the inverse operator was not supported, the system was only applicable to the simplest expressions.  More recently, several other solutions to different variants of this problem have been developed: The Formal Linear Algebra Methods Environment (FLAME) \cite{Gunnels:2001gi,Bientinesi:2005hu} is a methodology for the derivation of algorithmic variants for basic linear algebra operations such as factorizations and the solution of linear systems; Cl1ck \cite{FabregatTraver:2011km,FabregatTraver:2011gu} is an automated implementation of the FLAME methodology.  The goal of BTO BLAS is to generate C code for bandwidth bound operations, such as fused matrix-vector operations \cite{Siek:2008ij}.  In contrast to the linear algebra compiler CLAK \cite{fabregat-traver2013a}, which inspired the code generation approach presented here, Linnea makes use of the algebraic nature of the domain to remove redundancy during the derivation.  DxTer uses domain knowledge to optimize programs represented as dataflow graphs \cite{marker2012, marker2015}.  LGen targets basic linear algebra operations for small operand sizes, a regime in which BLAS and LAPACK do not perform very well, by directly generating vectorized C code \cite{spampinato2016}.  SLinGen \cite{Spampinato:2018tz} combines Cl1ck and LGen to generate code for more complex small-scale problems, but still requires that the input is described as a sequence of basic operations.  Similar approaches for code generation exist for related domains such as tensor contractions (TCE \cite{Baumgartner:2005dq}) and linear transforms (Spiral \cite{Franchetti:eq}, FFTW \cite{Frigo:2005cp}).

Our aim is to combine the advantages of existing approaches: The simplicity, and thus, productivity, of a high-level language, paired with performance that comes close to what is achieved manually by human experts.  In \cite{barthels2019arxiv}, we described an earlier version of Linnea that used a breadth-first search algorithm, and investigated the sequential performance of the generated code on 25 application problems. With this article, we introduce an entirely new generation approach, based on a best-first search algorithm, accompanied by a thorough experimental evaluation.

\section{Algorithm Generation}
\label{sec:synthesis}

The core idea behind Linnea is to rewrite the input problem while successively identifying parts that are computable by a sequence of one or more of the available kernels. In general, for a given input problem and cost function, Linnea generates many different sequences, which all compute the problem, but differ in terms of cost.  In order to efficiently store all generated sequences, we use a graph in which nodes represent the input problem at different stages of the computation, and edges are annotated with the kernels used to transition from one stage (node) to another.

This process starts with a single root node containing a symbolic expression that represents the input problem.
The generation process consists of two steps, which are repeated until termination.
1) In the first step, the input expression is rewritten in different ways, for example by making use of distributivity.
The different representations of a given expression are not stored explicitly; instead, a node only contains one
canonical representation, and it is rewritten when necessary. 2) In the second step, on each representation of the
expression, different algorithms are used to identify subexpressions that can be computed with one or more of the
available kernels. Whenever such an expression is found, a new successor of the parent node is constructed.
The edge from the parent to the new child node is annotated with the sequence of kernels, 
and the child contains the expression that is left to be computed.

\begin{figure}
\begin{tikzpicture}
	\node (n1) [tbox] {$b := (X^T X)^{-1} X^T y$};
	\node (n2b) [tbox, below right=of n1, xshift=-0.5cm] {$b := R^{-1} Q^T y$};
	\node (n3c) [tbox, below=of n2b, yshift=-0.1cm] {$b := v_8$};

	\node (n2a) [tbox, below left=of n1, xshift=0.5cm, yshift=0cm] {$b := M_1^{-1} X^T y$};
	\node (n3a) [tbox, below=of n2a] {$b := L^{-T} L^{-1} X^T y$};
	\node (n4a) [tbox, below left=of n3a, xshift=2cm, yshift=-0.7cm] {$b := v_3$};
	\node (n4b) [tbox, below right=of n3a, xshift=-2cm, yshift=-0.7cm] {$b := v_6$};
	
	\path[->] (n1) edge [barrow] node [above right] {$(Q, R) \leftarrow \text{qr}(X)$} (n2b);
	\path[->] (n2b) edge [barrow] node [right, xshift=0.3cm] {
	$\begin{aligned}
		v_7 &\leftarrow Q^T y\\
		v_8 &\leftarrow R^{-1} v_7
	\end{aligned}$
	} (n3c);
	
	\path[->] (n1) edge [barrow] node [above left] {$M_1 \leftarrow X^T X$} (n2a);
	\path[->] (n2a) edge [barrow] node [left] {$L \leftarrow \text{chol}(M_1)$} (n3a);
	\path[->] (n3a) edge [barrow] node [left, xshift=-0.2cm] {
    	$\begin{aligned}
    		M_2 &\leftarrow L^{-1} X^T \\
    		v_1 &\leftarrow M_2 y \\
    		v_3 &\leftarrow L^{-T} v_1
    	\end{aligned}$
    } (n4a);
	\path[->] (n3a) edge [barrow] node [right, xshift=0.3cm] {
    	$\begin{aligned}
    		v_4 &\leftarrow X^T y \\
    		v_5 &\leftarrow L^{-1} v_4 \\
    		v_6 &\leftarrow L^{-T} v_5
    	\end{aligned}$
	} (n4b);
\end{tikzpicture}	

\caption{An example of a search graph for the input $b := (X^T X)^{-1} X^T y$. This graph represents only a small part of search space that Linnea actually explores.}
\label{fig:graphexample}
\end{figure}
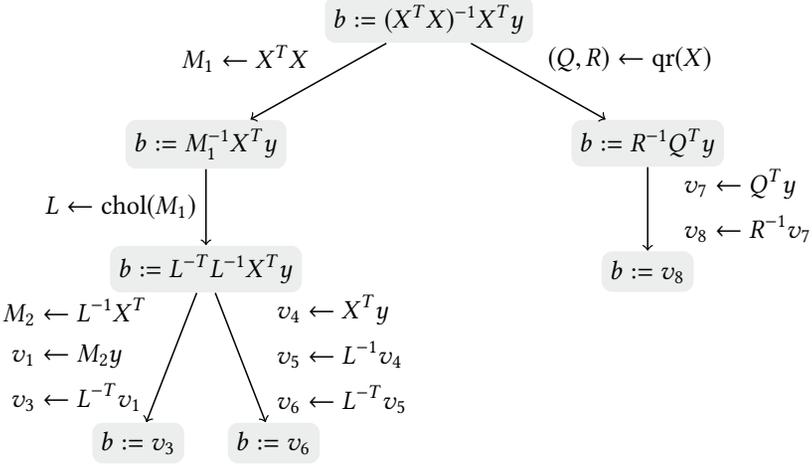

The two steps are then repeated on the new nodes, until one or more nodes with nothing left to compute are found or until the time limit is exceeded.
In practice, this process corresponds to the construction and traversal of a graph. An example of such a graph is shown in Fig.~\ref{fig:graphexample}.

Upon termination, the concatenation of all kernels along a path in the graph from the root to a leaf is a program that computes the input problem. In Sec.~\ref{sec:termination}, we discuss how termination is guaranteed. Given a function that assigns a cost to each kernel, the optimal program is found by searching for the shortest path in the graph from the root node to a leaf.

\subsection{The Algorithm}
\label{sec:algorithm}

In Linnea, the construction and traversal of the search graph is done with a best-first search algorithm. The rationale is to find a good, although potentially suboptimal solution as quickly as possible, to then use the cost of that solution to prune branches that cannot lead to a better solution. Over time, progressively better solutions are found. To guide the search towards good solutions, we use priorities to indicate which node to explore next. Priorities are non-negative integers, where smaller numbers indicate higher priority. In order to break ties and ensure that a first solution is found quickly, nodes are stored in a priority stack. This stack can be seen as a collection of stacks, one for each priority.
In a priority stack, the push operation corresponds to putting an element onto the internal stack of the corresponding priority. The pop operation instead takes an element from the top of the highest priority, non-empty stack.

The property that high priorities (i.e., small numbers) correspond to nodes that are likely to have a promising next successors is due to the following two facts:
1) The priority of a node is equal to the number of successors that have already been generated for this node, and 2) the ``next\_successor'' function, which returns a new successor of a node every time it is called, is designed to return the most promising successors first (this function is described in more detail in Sec.~\ref{sec:nextsuccessor}).
By using the number of current successors of a node as its priority, the algorithm effectively balances the number of successors of all nodes, that is,
it does not explore the $n + 1$th successor of any node before having explored the first $n$ successors of all nodes.
As a result, in contrast to depth-first search, the algorithm quickly goes back to the root node, instead of fully exploring the current branch first. The underlying idea is that the importance of a node does not depend on its position in the graph. While a node deep in the graph is closer to a solution, the decision that has the largest impact on the quality of this solution could have taken place already at the top of the graph.
As it is usually not necessary, and frequently also not practical to explore the full graph, we allow to specify an upper limit for the time spent on this search.

\begin{figure}
\hrule
\begin{lstlisting}
$G := (\{v_\text{input}\}, \varnothing) = (V, E)$ (*\hfill*) # graph initialization
$\text{best\_solution} := c_\infty$ (*\hfill*) # cost initialization
stack $:=$ PriorityStack() (*\hfill*) # stack initialization
stack.push(0, $v_\text{input}$) (*\hfill*) # the root node is added to the stack
while $\neg$stack.empty() and $\text{elapsed\_time} < t_\text{max}$:
	$(p, v) := \text{stack.pop}()$
	if $\text{cost}(v) > \text{best\_solution}$: (*\hfill*) # node is pruned
		continue (*\hfill*) # jump to line 5
	$v_\text{new} := \text{next\_successor}(v)$ (*\hfill*) # child creation
    $V := V \cup \{ v_\text{new} \}$
    $E := E \cup \{ (v, v_\text{new}) \}$
	if $\neg \text{is\_terminal}(v_\text{new})$:
		stack.push(0, $v_\text{new}$)
	else:
	    if $\text{cost}(v_\text{new}) < \text{best\_solution}$:
			$\text{best\_solution} := \text{cost}(v_\text{new})$
	stack.push($p+1$, $v$) (*\hfill*) # the current node is added back to the stack
\end{lstlisting}
\hrule
\caption{Pseudocode of the code generation algorithm.
}

\label{fig:pseudocodenm}
\end{figure}

The algorithm is shown in Fig.~\ref{fig:pseudocodenm}. The search graph is initialized with $v_\text{input}$ as root node in line 1; the variable ``best\_solution'' will hold the cost of the current best solution, and is initialized with infinity in line 2; the priority stack initially contains $v_\text{input}$ with priority $0$ (line 4). At every iteration of the while loop, a new successor is generated. To this end, in line 6 the node with the highest priority is taken from the stack. This operation returns both the node $v$, as well as its priority $p$. If the cost of $v$ (the cost of the path from the root node to $v$), is higher than the cost of the current best solution, then node $v$ is pruned (it cannot lead to a better solution), and the rest of the loop body is skipped (lines 7--8). If $v$ is not pruned, then its next successor, $v_\text{new}$, is generated in line 9; $\text{cost}(v_\text{new})$ is set to the sum of $\text{cost}(v)$ and the cost of the kernel(s) along the edge from $v$ to $v_\text{new}$.
Although not shown in the code, if $v_\text{new}$ does not exist because all successors were already explored, the rest of the loop body is skipped too. If $v_\text{new}$ is a terminal node, that is, there is nothing left to compute, ``best\_solution'' may have to be updated with $\text{cost}(v_\text{new})$ (lines 15--16); if $v_\text{new}$ is not terminal, in line 13 it is added to the stack with priority $0$. Finally, in line 17, the node $v$ is put back in the stack with priority $p+1$.  The loop terminates either when the stack is empty, or when the time limit is reached.

\subsection{Redundancy in the Derivation Graph}
\label{sec:redundancy}

\begin{figure}
\begin{tikzpicture}[node distance=1cm]

\node (n1) [tbox]  {$X := A(B+C+DE)$};

\node (n2a) [tbox, below left=of n1, xshift=1.3cm, yshift=-0.7cm]  {$X := M_3 + ADE$};

\node (n2b) [tbox, below right=of n1, xshift=-1.3cm, yshift=-0.7cm]  {$X := M_5 + ADE$};

\path[->] (n1) edge [barrow] node [left, xshift=-0.7cm] {
	$\begin{aligned}
		M_1 &\leftarrow AB \\
		M_2 &\leftarrow AC \\
		M_3 &\leftarrow M_1 + M_2
	\end{aligned}$
	} (n2a);

\path[->] (n1) edge [barrow] node [right, xshift=0.4cm] {
	$\begin{aligned}
		M_4 &\leftarrow B + C \\
		M_5 &\leftarrow A M_4
	\end{aligned}$
	} (n2b);

\end{tikzpicture}
\caption{Derivation graph with redundancy.}
\label{fig:dgredundancy}
\end{figure}
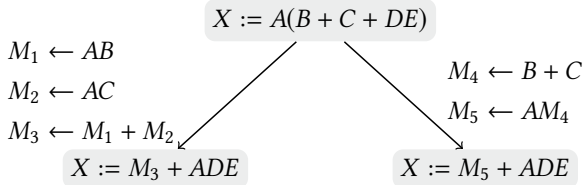

With large input expressions, it frequently happens that there is a lot of redundancy in the derivation graph. As an example, to compute the subexpressions $A(B+C)$ of $A(B+C+DE)$, the two different programs shown in Fig.~\ref{fig:dgredundancy} were constructed. As the generation process unfolds, both leaf nodes will be expanded, deriving the same programs for $ADE$ twice.
This phenomenon can be alleviated by taking advantage of the algebraic nature of the domain. In Fig.~\ref{fig:dgredundancy}, it is clear that $M_3$ and $M_5$ represent the same quantity because $AB+AC = A(B+C)$.\footnote{Ignoring differences due to floating-point arithmetic.} Thus, it is possible to merge the two branches and only do the derivation for $ADE$ once.

Our approach for detecting equivalent nodes and for merging branch\-es in the derivation graph consists of two parts: First, we define a normal form for expressions, that is, a unique representation for algebraically equivalent terms. Then, we make sure that irrespective of how a subexpression was computed, its result is always represented by the same, unique intermediate operand. In case of the graph in Fig.~\ref{fig:dgredundancy}, this would mean that the same intermediate for $AB+AC = A(B+C)$ is used in both leaves. When rewritten to their normal form, the equivalence of two expressions can simply be checked by a syntactic comparison.

\subsubsection{Normal Form for Expressions}

As a normal form for linear algebra expressions, both a \emph{sum of products} (e.g., $AB + AC$) and a \emph{product of sums} (e.g., $A(B+C)$) can be used; we opted for the sum of products.\footnote{The reason is that it is not obvious how to make the product of sums form unique. As an example, the expression $AC + AD + BC$ can be written both as $A(C+D) + BC$ and $(A+B)C+AD$.} Terms in sums are sorted according to an arbitrary total ordering on terms. The transposition and inversion operators are pushed down as far as possible: As examples, the normal form of $(AB)^{-1}$ and $(A+B)^T$ is $B^{-1} A^{-1}$ and $A^T+B^T$, respectively. Since expressions are converted between different representations during the derivation, the normal form does not influence the quality of the generated code.

Deciding whether or not two different representations represent the same element of an algebra is known as the \emph{word problem}, which in many cases is undecidable~\cite[pp. 59--60]{Baader:1999uu}. At least for some cases, this problem can be solved by a confluent and terminating term rewriting system, which can be obtained with  the Knuth-Bendix completion algorithm, or some of its extensions (for an overview, consider \cite{Dick:1991jh}). In practice, the merging of branch still works if some terms cannot correctly be identified as equivalent. This simply has the effect that some opportunities for merging will not be identified, so the optimization is less effective.

\subsubsection{Unique Intermediate Operands}

To ensure that the same intermediate operand is used for equivalent expressions, we make use of the normal form of expressions. The idea is to maintain a table of intermediate operands and the expressions they represent in the normal form. Whenever a kernel is used to compute part of an expression, we reconstruct the full expression that is computed by recursively replacing all intermediate operands. The resulting expression is then transformed to its normal form, and it is checked if there already is an intermediate operand for this expression in the table of intermediate operands.

\begin{example}

Let us assume we are given the input $X:= A(B+C+D)$. Initially, the table of intermediate operands, which is shown in Tab.~\ref{tab:intermediates}, is empty. The first partial program is found by rewriting this assignment as $X:= AB+AC+AD$ and computing
$$T_1 \leftarrow AB \qquad T_2 \leftarrow AC \qquad T_3 \leftarrow T_1 + T_2\text{,}$$
resulting in $X := T_4 + AD$. For the first two kernels, we simply add the intermediates $T_1$ and $T_2$, and the
corresponding expressions $AB$ and $AC$ to the table. For $T_1 + T_2$, we first use the table to replace the
intermediate operands $T_1$ and $T_2$ with the expressions they represent, resulting in $AB+AC$. As this expression is
already in normal form, we can simply check if there already is an entry for it in the table. Since at this point, there
is no entry for $AB+AC$ yet, $AB+AC$ is added to the table, and a new operand $T_3$ is created.
Alternatively, the same part of $X:= A(B+C+D)$ can be computed as
$$T_4 \leftarrow B+C \qquad T_3 \leftarrow A T_4\text{.}$$
For the kernel invocation $A T_4$, the intermediate operand is created by replacing $T_4$ by $B+C$, and then converting the resulting expression $A(B+C)$ to normal form, which in this case is $AB+AC$. Then, from the table, $T_3$ is retrieved. Tab.~\ref{tab:intermediates} shows the state of the table after deriving those two programs.\qed
\begin{table}
\caption{The table of intermediate operands after deriving two programs computing the subexpression $A(B+C)$ in $X:= A(B+C+D)$.}
\begin{tabular}{c c}
	\toprule
	intermediate & expression \\
	\midrule
	$T_1$ & $AB$ \\
	$T_2$ & $AC$ \\
	$T_3$ & $AB+AC$ \\
	$T_4$ & $B+C$ \\
	\bottomrule
\end{tabular}
\label{tab:intermediates}
\end{table}
\end{example}

\subsubsection{Merging Branches}

When merging branches, we implicitly assume that nodes do not have any state information such as the state of the registers, caches, or memory. This is a simplification that does not hold true in reality. However, without this assumption, it would not be possible to merge branches in the derivation graph. This optimization can drastically reduce the size of the derivation graph without reducing the size of the search space, thus making it possible to generate programs for larger input expressions.

\begin{figure}
\centering
\begin{tikzpicture}[node distance=1cm]

\coordinate (leftanchor) at (-2,0);
\coordinate (rightanchor) at (2,0);

\node (ln) at (leftanchor) [xshift=-0.5cm, yshift=1cm] {before:};

\node (ln1) at (leftanchor) [tbox]  {10};

\node (ln2) [tbox, below=of ln1, fill=rwthblue50]  {30};

\node (ln3) [tbox, below right=of ln2, xshift=-0.5cm, fill=rwthred50]  {44};

\node (ln4) [tbox, below left=of ln3, xshift=0.5cm, fill=rwthred50]  {37};

\node (ln5) [tbox, right=of ln1]  {5};

\node (ln6) [tbox, below=of ln5, fill=rwthblue50]  {20};

\path[->] (ln1) edge [barrow] node [left] {20} (ln2);
\path[->] (ln2) edge [bend left, barrow] node [right] {14} (ln3);
\path[->] (ln2) edge [bend right=40, barrow] node [left] {7} (ln4);
\path[->] (ln3) edge [bend left, barrow] node [right] {3} (ln4);
\path[->] (ln5) edge [barrow] node [right] {15} (ln6);

\draw (0.65,1.5) -- (0.65,-4.5);

\node (rn) at (rightanchor) [xshift=-0.5cm, yshift=1cm] {after:};

\node (rn1) at (rightanchor) [tbox]  {10};

\node (rn2) [tbox, below=of rn1, fill=rwthblue50]  {20};

\node (rn3) [tbox, below right=of rn2, xshift=-0.5cm, fill=rwthred50]  {34};

\node (rn4) [tbox, below left=of rn3, xshift=0.5cm, fill=rwthgreen50]  {27};

\node (rn5) [tbox, right=of rn1]  {5};

\path[->] (rn1) edge [barrow] node [left] {20} (rn2);
\path[->] (rn2) edge [bend left, barrow] node [right] {14} (rn3);
\path[->] (rn2) edge [bend right=40, barrow] node [left] {7} (rn4);
\path[->] (rn3) edge [bend left, barrow] node [right] {3} (rn4);
\path[->] (rn5) edge [bend left=20, barrow] node [right] {15} (rn2);

\end{tikzpicture}
\caption{An example of how through merging, pruned nodes need to be reactivated. Let us assume that the cost of the best solution is $32$, so initially, the two red nodes are pruned. When the blue nodes are merged, the cost of one of the pruned nodes decreases below $32$, so it may now lead to a new solution with a cost below $32$.}
\label{fig:mergingcost}
\end{figure}
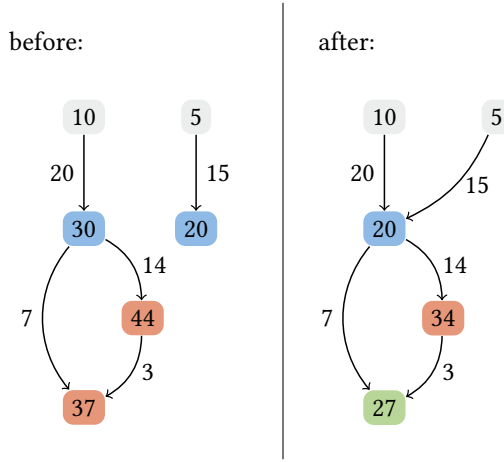

\subsubsection{Updated Algorithm}
Branch merging is an optimization that imposes some changes in the search algorithm; when a new node is generated, it
has to be checked whether or not it is equivalent to an already existing one. If there is an equivalent node, the new node is merged into the existing one. In addition, through merging, it is possible that the cost of a pruned node decreases, such that the node can again lead to a new, better solution. As a result, pruned nodes need to be reactivated. An example of how merging affects pruned nodes is shown in Fig.~\ref{fig:mergingcost}.

\section{Implementation}
\label{sec:implementation}

\subsection{Symbolic Expressions and Pattern Matching}

\begin{figure}
\begin{tikzpicture}[level distance=1.1cm, sibling distance=1cm]
	\node {$\times$}
		child {node {$A$}}
		child {node {$+$}
			child {node {$B$}}
			child {node {$C$}}
			child {node {$D$}}
		}
	;
\end{tikzpicture}
\caption{Expression tree for $A(B+C+D)$.}
\label{fig:exprtree}
\end{figure}
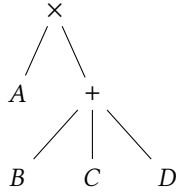

In Linnea, the input problem is represented as a symbolic expression, i.e., a tree-like algebraic data structure constructed from function symbols and constants that represent operands. As an example, the expression $A(B+C+D)$ is represented by the tree shown in Fig.~\ref{fig:exprtree}. Instead of using nested binary operations, and thus nested binary expression trees, associative operations such as multiplication and addition are flattened to n-ary operations.

Each available kernel is represented by a pattern, that is, a symbolic expression with variables. For instance, the \texttt{gemm} kernel for matrix-matrix multiplication (with $\alpha = \beta = 1$) is described by the pattern $X Y + Z$, where $X$, $Y$ and $Z$ are variables that match a single matrix. To identify where kernels can be applied, we use associative-commutative pattern matching \cite{Krebber:2017tp}. A search for the pattern $X Y + Z$ in an expression then yields all subexpressions that can be computed with the \texttt{gemm} kernel. Consider as an example the pattern $XY$, for the \texttt{gemm} kernel with $\alpha = 1$ and $\beta = 0$. In the expression $ABC$, two matches are found; $AB$ and $CD$.

Associative-commutative pattern matching makes it possible to specify for each operator if it is associative and/or commutative. The pattern matching algorithm automatically takes those properties into account. For instance, with the pattern $X^T +Y$ and the expression $A + B^T + C$, two matches are found: $B^T + A$ and $B^T + C$; since addition is commutative, these two matches are found irrespective of how the terms in $A + B^T + C$ ordered.

To make use of specialized kernels that exploit the properties of matrices, we use patterns with constraints on the variables. A pattern only matches if the constraints for all operands are satisfied.

We use the Python module MatchPy~\cite{krebber2017scipy, krebber2018joss}, which offers efficient many-to-one algorithms for associative-commutative pattern matching. For many-to-one matching, data structures similar to decision trees make it possible to use similarities between patterns to speed up matching~\cite{krebber2017pyhpc}.

\subsection{Matrix Properties}
\label{sec:properties}

Linnea's input format makes it possible to annotate matrices with properties. However, not only is it important to know the properties of the input matrices, it is at least equally important to know the properties of intermediate operands, as the computation unfolds. Consider for instance the generalized least squares problem $b := (X^T M^{-1} X)^{-1} X^T M^{-1} y$. Since $X$ has full rank and more rows than columns, and $M$ is symmetric positive definite, it can be inferred that $X^T M^{-1} X$ is symmetric positive definite, irrespective of how it is computed. This knowledge then allows one to solve the linear system $(\ldots)^{-1} X^T M^{-1} y$ with a Cholesky factorization, as opposed to the more expensive LU factorization.

To infer and propagate matrix properties, we encoded linear algebra knowledge into a set of inference rules such as
\begin{align*}
	\lotri{A} &\rightarrow \uptri{A^T}\\
	\diag{A} \land \diag{B} &\rightarrow \diag{AB}\\
	A = A^T &\rightarrow \sym{A}\text{,}
\end{align*}
where $A$ and $B$ are arbitrary matrix expressions.

\subsection{Factorizations}

In contrast to other languages and libraries, in the input, Linnea does not distinguish between the explicit matrix
inversion and the solution of a linear system. Whenever possible, inversion is avoided in favor of a linear system;
matrices are explicitly inverted only if this is unavoidable, for example in expressions such as $A^{-1} + B$. Even
though LAPACK offers kernels that encapsulate a factorization followed by a linear system solve (e.g., \texttt{gesv}),
Linnea ignores those kernels and applies factorizations directly. This is because the explicit factorization might enable other optimizations which are not possible when using a ``black box'' kernel such as \texttt{gesv}. As an example, consider again the generalized least squares problem $b := (X^T M^{-1} X)^{-1} X^T M^{-1} y$. This problem can be computed efficiently by applying the Cholesky factorization to $M$, resulting in $b := (X^T L^{-1} L^{-T} X)^{-1} X^T L^{-1} L^{-T} y$. In this expression, the subexpression $X^T L^{-1}$ or its transpose $L^{-T} X$ appears three times and only needs to be computed once. If either $X^T M^{-1}$ or $M^{-1} X$ were computed with a single kernel, this redundancy would not be exposed and exploited. Furthermore, the use of the Cholesky factorization allows to maintain the symmetry of $X^T M^{-1} X$.

Linnea uses the following factorizations: Cholesky, LU, QR, symmetric eigenvalue decomposition and singular value decomposition.
$LDL^T$ is currently not supported, because with the current LAPACK interface, it is not possible to separately access $L$ and $D$; they can only be used in kernels to directly solve linear systems or invert matrices. Factorizations are only applied to operands that appear inside of the inversion operation, and are not applied to triangular, diagonal and orthogonal operands.

\subsection{Termination}
\label{sec:termination}

Whether or not Linnea is able to find a solution for a given input problem depends on the set of kernels.  Trivially, to guarantee that a solution exists, it is sufficient to have one kernel for every supported operation.  In practice, Linnea uses a much larger set, including multiple kernels for the same operations that make use of different properties, as well as kernels that combine multiple operations.  With such a set of kernels, termination is guaranteed because every application of a kernel decreases the size of the input problem.  However, since Linnea also directly uses matrix factorizations, care has to be taken; repeatedly applying a matrix factorization and then undoing it by a matrix product can easily lead to infinite loops. In Linnea, such loops are avoided by labeling operands as factors and by requiring that for any given kernel call, there must be at least one operand that is not a factor.  For instance, in the expression $S^{-1} B$, the Cholesky factorization is applied to $S$, resulting in $(L^T L)^{-1} B$. To compute the resulting expression, first the inverse has to be distributed over $L^T L$, yielding $L^{-1} L^{-T} B$. Then, the linear system $M = L^{-T} B$ is solved, which is allowed because $B$ is not a factor,
and the remaining linear system $L^{-1} M$ can be solved too because $M$ is not a factor either.

\subsection{Rewriting Expressions}
\label{sec:rewriting}

In Sec.~\ref{sec:redundancy}, we discussed the conversion of expressions to normal form. In addition, to explore different, algebraically equivalent formulations of a problem, Linnea uses functions to rewrite expressions into alternative forms. Expressions in normal form are rewritten in several ways: Distributivity is used to convert expressions to products of sums. If possible, the inverse operator is pushed up, so $B^{-1} A^{-1}$ is also represented as $(AB)^{-1}$.

To explore an even larger set of alternatives, we developed an algorithm to detect common subexpressions of arbitrary length that takes into account identities such as $B^T A^T = (AB)^T$ and $B^{-1} A^{-1} = (AB)^{-1}$. As a result, even terms such as $A^{-1}B$ and $B^T A^{-T}$ are identified as a common subexpression. Since the use of a common subexpression does not necessarily lead to lower computational cost, Linnea also continues to operate on the unmodified expressions. Existing methods for the elimination of redundancy in code, such as common subexpression elimination, partial redundancy elimination, global value numbering \cite[Chap. 13]{Muchnick:1997wv}, are not able to consider algebraic identities.

In addition to those relatively general rewritings, we also encoded a small number of non-trivial rules that allow to
compute specific terms at a reduced cost. For instance, $X := A^T A + A^T B + B^T A$ becomes $Y := B + A/2$ and $X := A^T Y + Y^T A$. While such transformations are only applicable in special cases, thanks to efficient many-to-one pattern matching, Linnea can identify such cases with only minimal impact on the overall performance.

\subsection{Cost Function}
\label{sec:cost}

For most inputs, Linnea generates many alternative programs, all mathematically equivalent, but with different
performance signatures and numerical properties.
To discriminate programs and to choose one that satisfies constraints such as memory usage, a cost function is
necessary. This can either be an exact cost or an estimate. Such a function could take into account the number and the
cost of kernel invocations (e.g., the number of floating-point operations performed, the number of bytes moved), and
even the numerical stability of the program.

A cost function has to fulfill two requirements: 1) It has to be defined on any sequence of one or more kernels, and 2)
a total ordering has to be defined on the costs.
For some simple functions, such as the number of floating-point operations (FLOPs), both conditions
are satisfied. For many others, the first condition poses a challenge. For example, while the
efficiency of individual kernels can be (tediously) modeled \cite{iakymchuk2012,peise2012},
the efficiency of an arbitrary sequence of kernels is expensive to obtain via measurements and
cannot be accurately derived by simply combining that of the individual kernels \cite{Peise:2014fr}.
Similarly, incorporating numerical stability into a cost function is a challenging task:
It is not necessarily clear how to represent an error analysis by means of one or few numbers, it is still difficult to
derive stability analyses even for individual kernels, and the analysis for a sequence of kernels is not a direct
composition of the analyses of the kernels \cite{Bientinesi:2011bv,higham1996}.

As a cost function, Linnea presently uses the number of FLOPs. This function has the advantage that it is easy to determine, and for the targeted regime of mid-to-large scale operands, it is usually a good proxy for the execution time.  An evaluation of the effectiveness of the number of FLOPs as a cost function is carried out in Sec.~\ref{sec:kbest}.

For each kernel, Linnea knows a formula that computes the number of FLOPs performed from the sizes of the matched operands. As an example, for the \texttt{gemm} kernel---which computes $A B + C$ with $A \in \mathbb{R}^{m \times k}$ and $B \in \mathbb{R}^{k \times n}$---the formula is $2mnk$. Those formulas were either taken from \cite[pp. 336--337]{higham2008}, or inferred by hand. To find the path in the derivation graph with the lowest cost, we use a $K$ shortest paths algorithm \cite{Jimenez:1999cd}. In case of ties, an arbitrary path is selected.

\subsection{Constructive Algorithms}
\label{sec:constructive}

While pattern matching alone is sufficient to find all possible algorithms, it has the disadvantage of exploring the (potentially very large) search space exhaustively.
For specific types of subexpressions, however, an exhaustive search is not necessary.
As an example, in expressions with high computational intensity, different parenthesizations in sums of matrices do not significantly affect performance.
For products of multiple matrices, on the other hand, different parenthesizations can make a large difference, but there is no need to exhaustively generate all of them.
Instead, efficient algorithms exist that find the optimal solution in terms FLOPs to this so called matrix chain problem in polynomial \cite{godbole1973} and log-linear time \cite{hu1982, hu1984}.

For those subexpressions, to find a first solution quickly, and to increase the chances that this solution is relatively
good, Linnea uses specialized algorithms. For sums, we developed a simple greedy algorithm.  For products, we use the generalized matrix chain algorithm \cite{barthels2018cgo}, which finds the optimal parenthesization for matrix chains containing transposed and inverted matrices and considers matrix properties.

\subsection{Successor Generation}
\label{sec:nextsuccessor}

A crucial part of Linnea's search algorithm is the design of the next\_successor function.  As mentioned in Sec.~\ref{sec:algorithm}, for a given node, next\_successor has to return the most promising successors first.  Given the large number of optimizations that Linnea applies, there are many design decisions that determine the behavior of this function. Most of these decisions are based on heuristics that encode the expertise of linear algebra library developers. Examples follow. When considering the different representations of an expression, the product of sums form is used first; the underlying idea is that this representation decreases the number of expensive multiplications: While $AB+AC$ requires two matrix-matrix multiplications, $A(B+C)$ requires only one. Whenever possible, the constructive algorithms for sums and products are used first, because they quickly lead to a relatively good, first solution; factorizations, on the other hand, are only considered relatively late because there are many cases where it is possible to apply them, but not necessary to find a solution.  Since it is very challenging to predict which optimization is the most promising for a given expression, we favor ``variety'' over depth; e.g., instead of first replacing all common subexpressions and then proceeding to factorizations, we replace one common subexpression, followed by one factorization, and continue in a round-robin fashion.

\subsection{Code Generation}

A path in the derivation graph is only a symbolic representation of an algorithm; it still has to be translated to actual code. Most importantly, all operands are represented symbolically, with no notion of where and how they are stored in memory. During the code generation, operands are assigned to memory locations, and it is decided in which storage format they are stored.

Many BLAS and LAPACK kernels overwrite one of their input operands. As an example, the \texttt{gemm} kernel $\alpha A B + \beta C$ writes the result into the memory location containing $C$.
Since Linnea currently only generates straight-line code, it can easily be determined with a basic liveness analysis if an operand can be overwritten.
If this is not the case, the operand is copied.
At present, Linnea does not reorder kernel calls to avoid unnecessary copies.

Some kernels use specialized storage formats for matrices with properties. As an example, for a lower triangular matrix, only the lower, non-zero part is stored. Those storage formats have to be considered when generating code: While specialized kernels for triangular matrices only access the non-zero entries, a more general kernel would read from the entire memory location. Thus, it has to be ensured that operands are always in the correct storage format, if necessary by converting the storage format. Similar to overwriting, storage formats are not considered during the generation of algorithms. During the code generation, operands are converted to different storage formats when necessary. The output of an algorithm is always converted back to the full storage format.

\section{Experiments}
\label{sec:experiments}

To evaluate Linnea, we perform three different experiments.%
\footnote{The code for the experiments is available at
\url{https://github.com/HPAC/linnea/tree/master/experiments}.
} First, we assess the quality of the code generated by Linnea by comparing against
Julia\footnote{Version 1.3.0.},
Matlab\footnote{Version 2019b.},
Eigen\footnote{Version 3.3.7.},
and Armadillo\footnote{Version 9.800.x.}.
We then investigate the generation time with and without merging branches, and conclude by evaluating the quality of the
cost function. 

The measurements were taken on a dual socket Intel Xeon E5-2680 v3 with 12 cores each, a clock speed of 2.2 GHz and 64 GB of RAM. For all but Matlab, we linked against the Intel MKL implementation of BLAS and LAPACK (MKL 2019 initial release) \cite{mkldoc}; Matlab instead uses MKL 2018 update 3.
For the execution of generated code, all reported timings refer to the minimum of 20 repetitions, each on cold data, to avoid any caching effects. The generation time was obtained from one single repetition, and for all experiments we limited it to 30 minutes.

\paragraph{Test Problems}

We use two different sets of test problems, one consisting of expressions coming from applications, and a synthetic one.
The first set consists of a collection of 25 problems from real applications, from domains such as image and signal
processing, statistics, and regularization. A representative selection of those problems is shown in
Appendix~\ref{sec:exampleproblems}; in these problems, the operand sizes are selected to reflect realistic use cases.
The second set consists of 100 randomly generated linear algebra expressions, each consisting of a single assignment.
The number of operands is chosen uniformly between 4 and 7. Operand dimensions are chosen uniformly between 50 and 2000 in steps of 50. 
We set square operands to have a 75\% probability to have one of the following properties: diagonal, lower triangular, upper triangular, symmetric, or symmetric positive definite. To introduce realistic common subexpressions, some expressions contain patterns of the form $XX^T$ and $XMX^T$, where $X$ is a subexpression with up to two matrices, and $M$ is a symmetric matrix.

\subsection{Libraries and Languages}

\begin{table}
\caption{Input representations for the expression $A^{-1} B C^T$, where $A$ is SPD and $C$ is lower triangular. The letters ``n'' and ``r'' denote the naive and recommended implementation, respectively.
}
\begin{tabular}{lp{6.1cm}} \toprule
	Name & Implementation \\\midrule
	Julia n & \texttt{inv(A)*B*transpose(C)} \\
	Julia r & \texttt{(A$\backslash$B)*transpose(C)} \\
	Armadillo n & \texttt{arma::inv\_sympd(A)*B*(C).t()} \\
	Armadillo r & \texttt{arma::solve(A, B)*C.t()} \\
	Eigen n & \texttt{A.inverse()*B*C.transpose()} \\
	Eigen r & \texttt{A.llt().solve(B)*C.transpose()} \\
	Matlab n & \texttt{inv(A)*B*transpose(C)} \\
	Matlab r & \texttt{(A$\backslash$B)*transpose(C)} \\\bottomrule
\end{tabular}
\label{tab:implementations}
\end{table}

For each library and language, two different implementations are used: \emph{naive} and \emph{recommended}. The naive implementation is the one that comes closest to the mathematical description of the problem. It is also closest to the input to Linnea. As examples, in Tab.~\ref{tab:implementations} we provide the implementations of $A^{-1} B C^T$, where $A$ is symmetric positive definite and $C$ is lower triangular.
Since documentations almost always discourage the use of the inverse operator to solve linear systems, we instead use dedicated functions, e.g. \texttt{A$\backslash$B}, in the recommended implementations.
The different implementations are described below. 
\begin{description}
	\item[Julia] Properties are expressed via types. The naive implementation uses \texttt{inv()}, while the recommended one uses the $\slash$ and $\backslash$ operators.
	\item[Matlab] The naive implementation uses \texttt{inv()}, the rec\-om\-mend\-ed one the $\slash$ and $\backslash$ operators.
	\item[Eigen] In the recommended implementation, matrix properties are described with views. For linear systems, we select solvers based on properties.
	\item[Armadillo] In the naive implementation, specialized functions are used for the inversion of SPD and diagonal matrices. For \texttt{solve}, we use the \texttt{solve\_opts::fast} option to disable an expensive refinement. In addition, \texttt{trimatu} and \texttt{trimatl} are used for triangular matrices.
\end{description}

\subsection{Quality of Generated Code}

\newcommand{\markscaling}{0.7}
\newcommand{\markscalingsmall}{0.5}

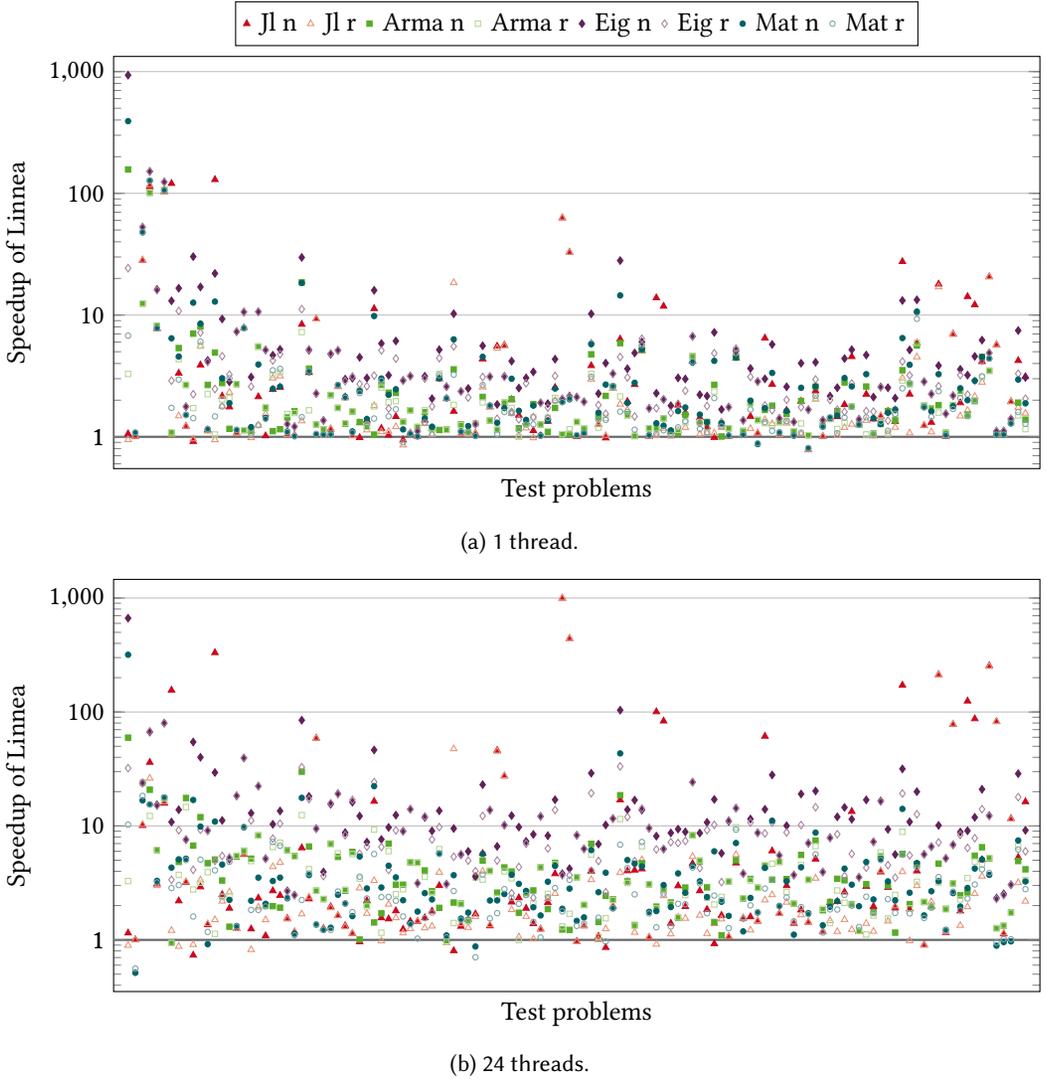
\begin{figure}[]
  \centering
  \begin{subfigure}{\textwidth}
  
    \pgfplotstableread[col sep=comma]{results_initial_submission/combined_t1_speedup.csv}\combinedspeedup
  
    \pgfplotstablesort[sort key=intensity, sort cmp=float <]\combinedspeedups{\combinedspeedup}
  
    \begin{tikzpicture}
      \begin{axis}[
          height=7cm,
          width=\columnwidth,
          xlabel={Test problems},
          ylabel={Speedup of Linnea},
          grid=major,
          ymode=log,
		xmin=-2,
		xmax=126,
        enlarge y limits=0.05,
		xmajorgrids=false,
		xtick=\empty,
		extra y ticks={1},
		extra y tick style={
			grid = major,
			grid style={
				color=rwthblack75,
				thick,
			}
		},
        legend style={
            column sep=0.2em,
            at={(0.5, 1.025)},
            anchor = south,
        },
        legend columns=8,
        log ticks with fixed point,
      ]
          \addplot[
              fill=rwthred,
              draw=rwthred,
              mark=triangle*,
              only marks,
              mark options={scale=\markscaling}
          ]
              table[
                  x expr=\coordindex,
                  y=naive_julia,
              ] {\combinedspeedups};
          \addlegendentry{Jl n};
          \addplot[
              fill=rwthred50,
              draw=rwthred50,
              mark=triangle,
              only marks,
              mark options={scale=\markscaling}
          ]
              table[
                  x expr=\coordindex,
                  y=recommended_julia,
              ] {\combinedspeedups};
          \addlegendentry{Jl r};
          \addplot[
              fill=rwthgreen,
              draw=rwthgreen,
              mark=square*,
              only marks,
              mark options={scale=\markscalingsmall}
          ]
              table[
                  x expr=\coordindex,
                  y=naive_armadillo,
              ] {\combinedspeedups};
          \addlegendentry{Arma n};
          \addplot[
              fill=rwthgreen50,
              draw=rwthgreen50,
              mark=square,
              only marks,
              mark options={scale=\markscalingsmall}
          ]
              table[
                  x expr=\coordindex,
                  y=recommended_armadillo,
              ] {\combinedspeedups};
          \addlegendentry{Arma r};
          \addplot[
              fill=rwthviolet,
              draw=rwthviolet,
              mark=diamond*,
              only marks,
              mark options={scale=\markscaling}
          ]
              table[
                  x expr=\coordindex,
                  y=naive_eigen,
              ] {\combinedspeedups};
          \addlegendentry{Eig n};
          \addplot[
              fill=rwthviolet50,
              draw=rwthviolet50,
              mark=diamond,
              only marks,
              mark options={scale=\markscaling}
          ]
              table[
                  x expr=\coordindex,
                  y=recommended_eigen,
              ] {\combinedspeedups};
          \addlegendentry{Eig r};
          \addplot[
              fill=rwthpetrol,
              draw=rwthpetrol,
              mark=*,
              only marks,
              mark options={scale=\markscalingsmall}
          ]
              table[
                  x expr=\coordindex,
                  y=naive_matlab,
              ] {\combinedspeedups};
          \addlegendentry{Mat n};
          \addplot[
              fill=rwthpetrol50,
              draw=rwthpetrol50,
              mark=o,
              only marks,
              mark options={scale=\markscalingsmall}
          ]
              table[
                  x expr=\coordindex,
                  y=recommended_matlab,
              ] {\combinedspeedups};
          \addlegendentry{Mat r};
      \end{axis}
    \end{tikzpicture}
    \caption{1 thread.}
    \label{fig:speedupst}
  \end{subfigure}\\[0.25cm]
  \begin{subfigure}{\textwidth}
  
    \pgfplotstableread[col sep=comma]{results_initial_submission/combined_t24_speedup.csv}\combinedspeedup
  
    \pgfplotstablesort[sort key=intensity, sort cmp=float <]\combinedspeedups{\combinedspeedup}
    
    \begin{tikzpicture}
      \begin{axis}[
        height=7cm,
        width=\columnwidth,
        xlabel={Test problems},
        ylabel={Speedup of Linnea},
        grid=major,
        ymode=log,
		xmin=-2,
		xmax=126,
        enlarge y limits=0.05,
		xmajorgrids=false,
		xtick=\empty,
		extra y ticks={1},
		extra y tick style={
          grid = major,
          grid style={
            color=rwthblack75,
            thick,
          }
		},
        legend style={
          column sep=0.2em,
          at={(0.5,-0.1)},
          anchor = north,
        },
        legend columns=8,
        log ticks with fixed point,
      ]
          \addplot[
              fill=rwthred,
              draw=rwthred,
              mark=triangle*,
              only marks,
              mark options={scale=\markscaling}
          ]
              table[
                  x expr=\coordindex,
                  y=naive_julia,
              ] {\combinedspeedups};
          \addlegendentry{Jl n};
          \addplot[
              fill=rwthred50,
              draw=rwthred50,
              mark=triangle,
              only marks,
              mark options={scale=\markscaling}
          ]
              table[
                  x expr=\coordindex,
                  y=recommended_julia,
              ] {\combinedspeedups};
          \addlegendentry{Jl r};
          \addplot[
              fill=rwthgreen,
              draw=rwthgreen,
              mark=square*,
              only marks,
              mark options={scale=\markscalingsmall}
          ]
              table[
                  x expr=\coordindex,
                  y=naive_armadillo,
              ] {\combinedspeedups};
          \addlegendentry{Arma n};
          \addplot[
              fill=rwthgreen50,
              draw=rwthgreen50,
              mark=square,
              only marks,
              mark options={scale=\markscalingsmall}
          ]
              table[
                  x expr=\coordindex,
                  y=recommended_armadillo,
              ] {\combinedspeedups};
          \addlegendentry{Arma r};
          \addplot[
              fill=rwthviolet,
              draw=rwthviolet,
              mark=diamond*,
              only marks,
              mark options={scale=\markscaling}
          ]
              table[
                  x expr=\coordindex,
                  y=naive_eigen,
              ] {\combinedspeedups};
          \addlegendentry{Eig n};
          \addplot[
              fill=rwthviolet50,
              draw=rwthviolet50,
              mark=diamond,
              only marks,
              mark options={scale=\markscaling}
          ]
              table[
                  x expr=\coordindex,
                  y=recommended_eigen,
              ] {\combinedspeedups};
          \addlegendentry{Eig r};
          \addplot[
              fill=rwthpetrol,
              draw=rwthpetrol,
              mark=*,
              only marks,
              mark options={scale=\markscalingsmall}
          ]
              table[
                  x expr=\coordindex,
                  y=naive_matlab,
              ] {\combinedspeedups};
          \addlegendentry{Mat n};
          \addplot[
              fill=rwthpetrol50,
              draw=rwthpetrol50,
              mark=o,
              only marks,
              mark options={scale=\markscalingsmall}
          ]
              table[
                  x expr=\coordindex,
                  y=recommended_matlab,
              ] {\combinedspeedups};
          \addlegendentry{Mat r};
          \legend{}; 
      \end{axis}
    \end{tikzpicture}
    \caption{24 threads.}
    \label{fig:speedupmt}
  \end{subfigure}
  \caption{Speedup of Linnea over four reference languages and libraries for 125 test problems. The test problems are
    sorted by computational intensity, increasing from left to right.}
  \label{fig:speedup}
\end{figure}

In Fig.~\ref{fig:speedup}, we present the speedups of the code generated by Linnea over other languages and libraries for both the random and application test cases.
For one and 24 threads, the code generated by Linnea is the fastest in 91\% and 82\% of the cases, respectively.
If not the fastest, the code is at most 1.3\su{} and 1.9\su{} slower than the other languages and libraries.
To understand where the speedups for the code generated by Linnea come from, we discuss the details of few exemplary test problems.

\paragraph{Distributivity}

The assignments
\begin{align*}
	H^\dag &:= H^T ( H H^T )^{-1} \text{,} \\
	y_k &:= H^\dag y + ( I_n - H^\dag H ) x_k\text{,}
\end{align*}
which are part of an image restoration application \cite{Tirer:2017uv}, illustrate well
how distributivity might affect performance.
Due to the matrix-matrix product $H^\dag H$, the computation of $y_k$ based on the original formulation of the problem
leads to $\mathcal{O}(n^3)$ FLOPs. Instead, for $y_k$ Linnea finds the solution
\begin{align*}
	v_\text{tmp} &:= - H x_k + y \\
	y_k &:= H^\dag v_\text{tmp} + x_k,
\end{align*}
which only uses matrix-vector products (\texttt{gemv}), and requires $\mathcal{O}(n^2)$ FLOPs.
This solution is obtained in two steps: First, $H^\dag y + ( I_n - H^\dag H ) x_k$ is converted to Linnea's normal form, returning $H^\dag y + x_k - H^\dag H x_k$; then, by factoring out $H^\dag$, the expression is written back as product of sums, resulting in $H^\dag (y - H x_k) + x_k$, which can be computed with two calls to \texttt{gemv}.
Here, this optimization yields speedups between 4.1\su{} (Matlab naive) and 6.7\su{} (Eigen recommended) with respect to the other languages and libraries for one thread, and speedups between 4.3\su{} (Matlab recommended) and 24\su{} (Eigen recommended) for 24 threads.

\paragraph{Associativity}
With the exception of Armadillo, none of the languages and libraries we compare with consider the matrix chain problem. Instead, products are always computed from left to right. The synthetic test case $X := M_1 M_1^T (M_2 + M_3) M_4 v_5 v_6^T$ is a good example to illustrate the importance of making use of associativity in products. The operands have the following dimensions: $M_1 \in \mathbb{R}^{150 \times 450}$, $M_2, M_3 \in \mathbb{R}^{150 \times 900}$, $M_4 \in \mathbb{R}^{900 \times 100}$, $v_5 \in \mathbb{R}^{100}$, and $v_6 \in \mathbb{R}^{150}$. All matrices are full.
Not only does Linnea successfully avoid any matrix-matrix products in the evaluation of this problem, surprisingly Linnea even finds a solution that avoids the sum $M_2 + M_3$. As a first step, the matrix-vector product $z_1 := M_4 v_5$ is computed. Then, Linnea rewrites the resulting $X := M_1 M_1^T (M_2 + M_3) z_1 v_6^T$ as $X := M_1 M_1^T M_2 z_1 v_6^T + M_1 M_1^T M_3 z_1 v_6^T$, where a second matrix-vector product $z_2 := M_3 z_1$ is computed. 
The resulting expression is rewritten again to $X := M_1 M_1^T (M_2 z_1 + z_2) v_6^T$, which is now computed as a sequence of three more matrix-vector products and one outer product:
\begin{align*}
  z_3 &:= M_2 z_1 + z_2 \\
  z_4 &:= M_1^T z_3 \\
  z_5 &:= M_1 z_4 \\
  X &:= z_5 v_6^T
\end{align*}
Despite the rather small operand sizes, the speedups for this test case are between 7.7\su{} and 16\su{} with one thread, and between 3.0\su{} and 15\su{} with 24 threads.

\paragraph{Common Subexpressions}
Expressions arising in application frequently exhibit common subexpressions; one such example is given by the assignment $$B_1 := \frac{1}{\lambda_1} (I_n - A^T W_1 (\lambda_1 I_l + W_1^T A A^T W_1)^{-1} W_1^T A )\text{,}$$ which is used in the solution of large least-squares problems \cite{Chung:2017ws}.
Linnea successfully identifies that the term $W_1^T A$ (or its transposed form $(A^T W_1)^T$) appears four times, and computes it only once. In this example, these savings lead to speedups between 5.1\su{} and 6.4\su{} with one thread, and between 4.2\su{} and 14\su{} with 24 threads.

\paragraph{Properties}
Many matrix operations can be sped up by taking advantage of matrix properties. As an example, here we discuss the evaluation of the assignment $x := (A^T A + \alpha^2 I)^{-1} A^T b$, a least-squares problem with Tikhonov regularization \cite{Golub:2006hl}, where matrix $A$ is of size $3000 \times 200$ and has full rank. Since $A$ has more rows than columns and is full rank, Linnea is able to infer that $A^T A$ is not only symmetric, but also positive definite (SPD). Similarly, Linnea infers that $\alpha^2 I$ is SPD because 1) the identity matrix is SPD, 2) $\alpha^2$ is positive and 3) a SPD matrix scaled by a positive factor is still SPD. Since the sum of two SPD matrices is SPD, $A^T A + \alpha^2 I$ is identified as SPD. As a result, the Cholesky factorization is used to solve the linear system. If $A^T A + \alpha^2 I$ had not been identified as SPD, a more expensive factorization such as LU had to be used. Finally, since Linnea infers properties based on the annotations of the input matrices, no property checks have to be performed at runtime; if the input matrices have the specified properties, all inferred properties hold. Altogether, the code generated for this assignment is between 1.2\su{}  and 5.2\su{} faster than the other languages and libraries with one thread, and 1.9\su{} and 14\su{} faster with 24 threads.

\subsubsection*{Epilog}

In general, the speedups of Linnea depend both on the potential for optimization in a given problem, as well as on the similarity of the default evaluation strategy in each language and library to the optimal one.

In case of problem \ref{exprob:tikhonov} for example, with one thread, the code generated by Linnea is 3.4\su{} faster than the recommended Armadillo implementation, but only 1.2\su{} faster than the naive implementation. The reason is that for this problem, the parenthesization has the largest influence on the execution time. While Armadillo does solve a simplified version of the matrix chain problem, the \texttt{solve} function used in the recommended implementation (see Tab.~\ref{tab:implementations}) effectively introduces a fixed parenthesization.
Due to the explicit inversion in the naive implementation, there is no such fixed parenthesization, so Armadillo is able to find a solution which is very similar to that generated by Linnea.

For problem \ref{exprob:triinv}, which is the loop body of a blocked algorithm for the inversion of a triangular matrix, there is a large spread between the speedups: The recommended Julia and Matlab solutions are respectively around 1.4\su{} and 1.5\su{} slower than Linnea, while the naive Matlab, Armadillo and Eigen implementations are respectively 18\su{}, 19\su{} and 30\su{} slower (one thread).
In this case, the large spread is mostly caused by a combination of the interface the different systems offer, and how they utilize properties.
Neither Armadillo nor Eigen have functions to solve linear systems of the form $A B^{-1}$, with the inverted matrix on the right-hand side. Thus, even in the recommended solution, for $X_{10} := L_{10} L_{00}^{-1}$, explicit inversion is used instead. Armadillo and Eigen are not able to identify that $L_{00}$ is lower triangular and instead use an algorithm for the inversion of a general matrix, leading to a significant loss in performance, while Julia and Matlab correctly use the \texttt{trsm} kernel.

For expression \ref{exprob:randinv}, all solutions have very similar execution times; the speedups of Linnea are between 1.3\su{} and 1.9\su{} with one thread. The cost of computing this problem is dominated by the cost of computing the value of $X_{k+1}$, for which the solution found by all other languages and libraries is almost identical to the solution found by Linnea. While Linnea is able to save some FLOPs in the computation of $\Lambda$, those savings are negligible for the evaluation of the entire problem. With 24 threads, there is a larger spread, with speedups ranging from 1.0\su{} to 12\su{}. This spread is likely caused by the differences in how well the operations not supported by BLAS and LAPACK are parallelized.

\subsection{Generation Time and Merging}

The search algorithm gives Linnea flexibility: A potentially suboptimal solution can be found quickly, and better ones
can be found if more time is invested. In the following, we distinguish between 1) the time needed for the construction
of the graph, and 2) the time needed to retrieve the best solution from the graph and to translate this into code.
In this section, we use $A_\text{1st}$ to refer to the solution that is found first, and $A_\text{30min}$ for the best solution, according to the cost function, that can be found within 30 minutes.

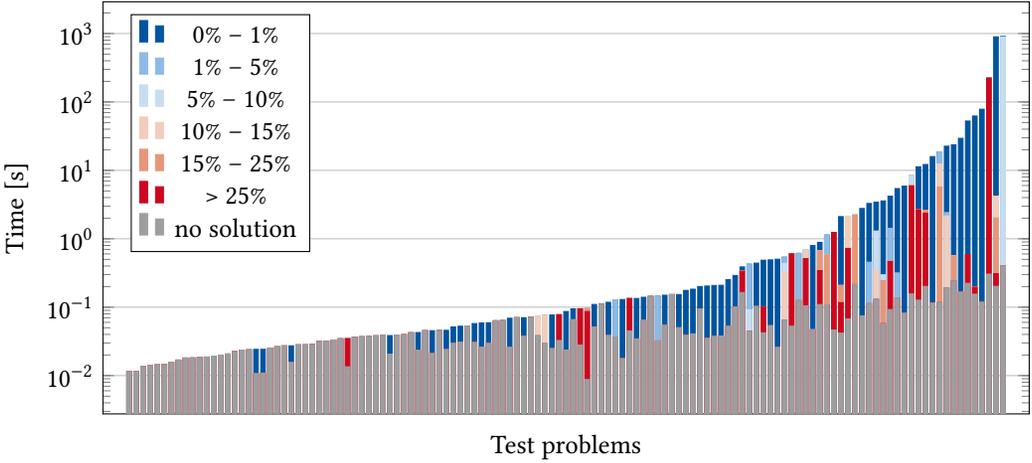
\begin{figure}[]
    \centering
    \begin{tikzpicture}[]
        \begin{axis}[
            height=7cm,
            width=\textwidth,
            xlabel={Test problems},
            ylabel={Time [s]},
            grid=major,
            ybar=0pt,
            bar width=0.65,
            bar shift=0,
            xtick=\empty,
            ymode=log,
            log origin=infty,
            enlarge x limits=0.03,
        	extra x tick style={
        		draw=none,
        		grid=none,
        		font=\normalsize,
        		text height=1.5ex,
        		anchor=north
        	},
            legend style={
            	column sep=0.2em,
            },
			legend pos=north west,
            legend columns=1,
        ]
            \addplot[
            	fill=rwthblue,
            	draw=rwthblue
            ]
                table[
                    x expr=\coordindex,
                    y=best_solution_time,
                ] {\generationmergingsorted};
            \addlegendentry{0\% -- 1\%};
            \addplot[
            	fill=rwthblue50,
            	draw=rwthblue50
            ]
                table[
                    x expr=\coordindex,
                    y=time_to_1pc,
                ] {\generationmergingsorted};
            \addlegendentry{1\% -- 5\%};
            \addplot[
            	fill=rwthblue25,
            	draw=rwthblue25
            ]
                table[
                    x expr=\coordindex,
                    y=time_to_5pc,
                ] {\generationmergingsorted};
            \addlegendentry{5\% -- 10\%};
            \addplot[
            	fill=rwthred25,
            	draw=rwthred25
            ]
                table[
                    x expr=\coordindex,
                    y=time_to_10pc,
                ] {\generationmergingsorted};
            \addlegendentry{10\% -- 15\%};
            \addplot[
            	fill=rwthred50,
            	draw=rwthred50
            ]
                table[
                    x expr=\coordindex,
                    y=time_to_15pc,
                ] {\generationmergingsorted};
            \addlegendentry{15\% -- 25\%};
            \addplot[
            	fill=rwthred,
            	draw=rwthred
            ]
                table[
                    x expr=\coordindex,
                    y=time_to_25pc,
                ] {\generationmergingsorted};
            \addlegendentry{> 25\%};
            \addplot[
            	fill=rwthblack50,
            	draw=rwthblack50
            ]
                table[
                    x expr=\coordindex,
                    y=first_solution_time,
                ] {\generationmergingsorted};
            \addlegendentry{no solution};
        \end{axis}
    \end{tikzpicture}
    \caption{Code generation time in Linnea. The height of the bars indicates the time to find $A_\text{30min}$. The colors indicate the cost of the current best solution at any given time, relative to the cost of $A_\text{30min}$. The relative cost is given as the overhead over the best solution, in percent. Gray indicates that no solution has been found yet.}
    \label{fig:generationtime}
\end{figure}

Fig.~\ref{fig:generationtime} reports for all 125 test problems the graph construction time and the quality of the different solutions found over time.
For all problems, $A_\text{1st}$ is found in less than one second; for 79\% of the problems, also $A_\text{30min}$ is found in less than one second.
In only 3 cases, the optimal solution is found after more than two minutes.
In terms of FLOPs, $A_\text{1st}$ is within 25\% of $A_\text{30min}$ for 83\% of the test problems, and within 1\% in 70\% of the cases.

The average time to retrieve the best algorithm from the graph, generate the code, and write it to a file is 0.1 second; in the worst case, it is 0.6 seconds.

\subsubsection{Impact of Merging Branches}
As discussed in Sec.~\ref{sec:redundancy}, in order to reduce the size of the search graph and thus speed up program
generation, redundant branches in the derivation graph are merged. To evaluate the impact of this optimization, we
performed the code generation with this optimization enabled and disabled. Since merging branches only
reduces redundancy without eliminating any solutions, given sufficient time, the same solutions will be found. As the
search graph initially contains very little redundancy, the time to find $A_\text{1st}$ is mostly unaffected by the merging. There are, however, notable differences in the time to find $A_\text{30min}$, especially for those problems where the best solution is not found within a few seconds. Without merging, there are 14 test problems for which the best solution found with merging is not found within 30 minutes. In 32 cases, it takes more than twice as long to find $A_\text{30min}$, including 11 cases where it takes at least 10 times longer.

\pgfplotstableread[col sep=comma]{results_initial_submission/combined_generation_merging.csv}\generationmerging
\pgfplotstablesort[sort key=best_solution_time,sort cmp=float <]\generationmergingsorted{\generationmerging}

\subsection{Quality of the Cost Function}
\label{sec:kbest}
  
\pgfplotstableread[col sep=comma]{results_initial_submission/combined_t24_k_best_stats_min_time.csv}\kbest

\pgfplotstablesort[sort key=intensity, sort cmp=float <]\kbests{\kbest}

\begin{figure}[]
    \centering
    \begin{tikzpicture}
        \begin{axis}[
            height=7cm,
            width=\columnwidth,
            xlabel={Test problems},
            ylabel={Speedup --- Relative Cost},
            grid=major,
			xmin=-2,
			xmax=126,
            enlarge y limits=0.05,
			xmajorgrids=false,
			xtick=\empty,
			extra y ticks={1},
			extra y tick style={
				grid = major,
				grid style={
					color=rwthblack75,
					thick,
				}
			},
            legend style={
                column sep=0.2em,
            },
			legend pos=north east,
            legend columns=4,
            log ticks with fixed point,
        ]
            \addplot[
                fill=rwthred,
                draw=rwthred,
                mark=*,
                only marks,
                mark options={scale=\markscalingsmall}
            ]
                table[
                    x expr=\coordindex,
                    y=speedup,
                ] {\kbests};
            \addlegendentry{Speedup};
            \addplot[
                fill=rwthblack50,
                draw=rwthblack50,
                ybar,
                ybar legend,
                bar width=0.65,
            ]
                table[
                    x expr=\coordindex,
                    y=cost_rel,
                ] {\kbests};
            \addlegendentry{Relative Cost};
        \end{axis}
    \end{tikzpicture}
    \caption{Comparison between $A_\text{time}$ and  $A_\text{FLOPs}$ in terms of execution time (dots) and FLOP count (bars) for 24 threads.
    The test problems are sorted by computational intensity, increasing from left to right.}
    \label{fig:kbest}
\end{figure}
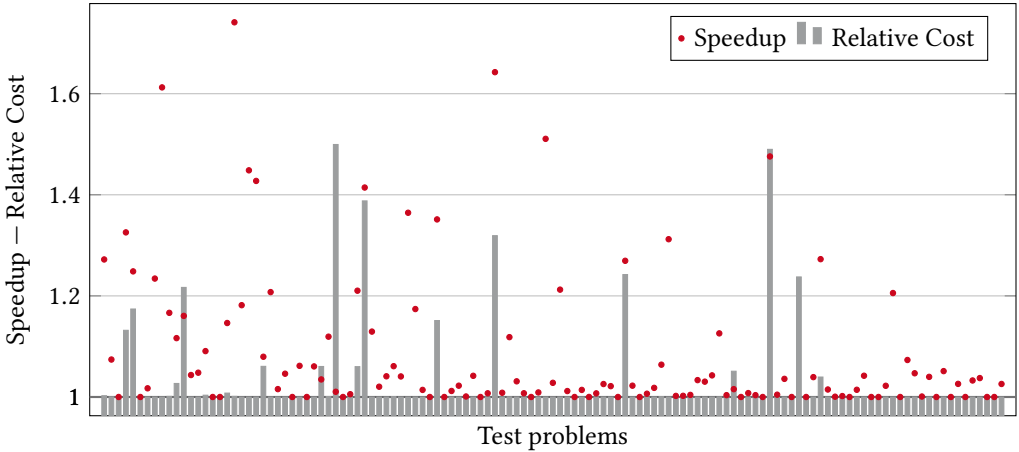

As a cost function, Linnea uses the number of FLOPs. To assess the accuracy of this function, we modified the pruning (at line 7 in the algorithm in Fig.~\ref{fig:pseudocodenm}) such that all algorithms with a cost of up to 1.5\su{} of the best solution are generated and run the best 100 of those algorithms per test problem.
In the following, we use $A_\text{FLOPs}$ to denote the algorithm that minimizes the cost function, and $A_\text{time}$ for the algorithm that is actually the fastest among all candidates.
For each test problem, we compare the number of FLOPs and the execution time of those two algorithms.
The results for 24 threads are shown in Fig.~\ref{fig:kbest}. In 109 cases, $A_\text{time}$ performs at most 1\% more FLOPs than $A_\text{FLOPs}$; there are only few cases where more FLOPs lead to a significantly lower execution time. The speedup of $A_\text{time}$ over $A_\text{FLOPs}$ is below 1.5\su{} in 120 cases. In the single-threaded case, the speedup is always below 1.1\su{}, and in all cases the relative cost is below 1.01\su{}.
It can be concluded that for the kind of problems that Linnea solves, the number of FLOPs is often a good indicator for
execution time and never entorely unreliable. This is especially true when the code is executed with one thread.
Most of the cases where the cost function is inaccurate are a result of not considering the efficiency of the kernels. Two examples where the cost function is particularly off follow.

As a first example, we consider the randomly generated test problem
\[X := M_1 (M_2^T M_3 M_4)^{-1} M_5,\] with $M_1 \in \mathbb{R}^{650 \times 1250}$, $M_2$ and $M_3 \in \mathbb{R}^{1700 \times 1250}$, $M_4 \in \mathbb{R}^{1250 \times 1250}$, and $M_5 \in \mathbb{R}^{1250 \times 1550}$; all matrices are full.
As a first step, both $A_\text{time}$ and $A_\text{FLOPs}$ compute $Z_1 := M_2^T M_3$, yielding $X := M_1 (Z_1 M_4)^{-1} M_5$.
At this point, in $A_\text{FLOPs}$ the LU factorization is applied to both $Z_1$ and $M_4$. After distributing the inverse, the problem becomes the matrix product $X := M_1 U_2^{-1} L_2^{-1} P_2 U_1^{-1} L_1^{-1} P_1 M_5$. The computation of this product, which is done from left to right, involves four calls to the \texttt{trsm} kernel for the solution of triangular linear systems. $A_\text{time}$ only uses one LU factorization and two triangular solves, but performs one additional matrix-matrix product: Instead of factoring $Z_1$ and $M_4$, those two matrices are multiplied together. After applying the LU factorization to the result of this product and distributing the inverse, $X := M_1 U_3^{-1} L_3^{-1} P_3 M_5$ is obtained. This product is again computed from left to right.  This algorithm requires about 4\% more FLOPs, but when executed with 24 threads is 27\% faster than the algorithm with the minimum number of FLOPs. While both the product $Z_1 M_4$, with $Z_1, M_4 \in \mathbb{R}^{1250 \times 1250}$, as well as one LU factorization ($1250 \times 1250$) and two triangular solves (with operands of size $650 \times 1250$ and $1250 \times 1250$) require almost the same number of FLOPs, the matrix-matrix product achieves a higher efficiency and is thus faster.




As a second example, we look at the randomly generated problem
\[X := M_1 M_2^T + M_3 M_3^T + M_4^T + M_5^T,\] 
with $M_1$ and $M_2 \in \mathbb{R}^{1100 \times 1800}$
$M_3 \in \mathbb{R}^{1100 \times 1150}$, and
$M_4$ as well as $M_5 \in \mathbb{R}^{1100 \times 1100}$. Matrices $M_4$ and $M_5$ are upper triangular, all others are full.
In $A_\text{FLOPs}$, the product $M_3 M_3^T$ is computed with the \texttt{syrk} kernel that makes use of symmetry. Since only half of the output matrix is stored, a storage format conversion is necessary to use this matrix in the following computations. In $A_\text{time}$, the same product is computed with a call to \texttt{gemm}. While this choice requires more FLOPs, it makes the storage format conversion unnecessary. The resulting algorithm performs 24\% more FLOPs, but is around 27\% percent faster with 24 threads. Again, this difference is caused by the higher parallel efficiency of \texttt{gemm} compared to \texttt{syrk} for matrices of the same size, but also by the storage format conversion.

The two examples discussed above are exceptions; for most of our test cases, the number of FLOPs is quite an accurate cost function.

\section{Conclusion and Future Work}
\label{sec:conclusion}

We presented Linnea, a code generator that translates a high-level linear algebra
problem into an efficient sequence of high-per\-for\-mance kernels. In contrast to other languages and libraries, Linnea uses domain knowledge such as associativity, commutativity, distributivity and matrix properties to derive efficient algorithms. Our experiments on randomly generated and application problems indicate that Linnea almost always outperforms all the current state-of-the-art tools. Linnea is also flexible, in that it can quickly return a good, but potentially not optimal solution, or invest more time into finding better solutions.

In the future, we aim to integrate the expected efficiency and scalability of kernels into the cost function. In addition, we plan to investigate different methods of parallelization, such as algorithms by blocks, offloading to accelerators, and the concurrent execution of kernels, as well as the extension to sparse and complex linear algebra, matrix functions, operands with block structure, multi-dimensional objects, and symbolic operand sizes.

\begin{acks}
Financial support from the Deutsche Forschungsgemeinschaft (German Research
Foundation) through grants GSC 111 and IRTG 2379 is gratefully
acknowledged. We thank Jan Vitek and Marcin Copik for their help.
\end{acks}

\bibliographystyle{ACM-Reference-Format}
\bibliography{../shared_tex_files/PhD_bibliography,../shared_tex_files/PhD_bibliography_papers,../shared_tex_files/my_publications}


\begin{thebibliography}{57}


\ifx \showCODEN    \undefined \def \showCODEN     #1{\unskip}     \fi
\ifx \showDOI      \undefined \def \showDOI       #1{#1}\fi
\ifx \showISBNx    \undefined \def \showISBNx     #1{\unskip}     \fi
\ifx \showISBNxiii \undefined \def \showISBNxiii  #1{\unskip}     \fi
\ifx \showISSN     \undefined \def \showISSN      #1{\unskip}     \fi
\ifx \showLCCN     \undefined \def \showLCCN      #1{\unskip}     \fi
\ifx \shownote     \undefined \def \shownote      #1{#1}          \fi
\ifx \showarticletitle \undefined \def \showarticletitle #1{#1}   \fi
\ifx \showURL      \undefined \def \showURL       {\relax}        \fi
\providecommand\bibfield[2]{#2}
\providecommand\bibinfo[2]{#2}
\providecommand\natexlab[1]{#1}
\providecommand\showeprint[2][]{arXiv:#2}

\bibitem[\protect\citeauthoryear{Aho, Ganapathi, and Tjiang}{Aho
  et~al\mbox{.}}{1989}]%
        {aho1989}
\bibfield{author}{\bibinfo{person}{Alfred~V. Aho}, \bibinfo{person}{Mahadevan
  Ganapathi}, {and} \bibinfo{person}{Steven W.~K. Tjiang}.}
  \bibinfo{year}{1989}\natexlab{}.
\newblock \showarticletitle{{C}ode {G}eneration {U}sing {T}ree {M}atching and
  {D}ynamic {P}rogramming}.
\newblock \bibinfo{journal}{\emph{ACM Trans. Program. Lang. Syst.}}
  \bibinfo{volume}{11}, \bibinfo{number}{4} (\bibinfo{date}{Oct.}
  \bibinfo{year}{1989}), \bibinfo{pages}{491--516}.
\newblock
\showISSN{0164-0925}
\urldef\tempurl%
\url{https://doi.org/10.1145/69558.75700}
\showDOI{\tempurl}


\bibitem[\protect\citeauthoryear{Aho and Johnson}{Aho and Johnson}{1976}]%
        {Aho:1976ga}
\bibfield{author}{\bibinfo{person}{Alfred~V. Aho} {and}
  \bibinfo{person}{Stephen~C. Johnson}.} \bibinfo{year}{1976}\natexlab{}.
\newblock \showarticletitle{{Optimal Code Generation for Expression Trees}}.
\newblock \bibinfo{journal}{\emph{Journal of the ACM (JACM)}}
  \bibinfo{volume}{23}, \bibinfo{number}{3} (\bibinfo{date}{July}
  \bibinfo{year}{1976}), \bibinfo{pages}{488--501}.
\newblock


\bibitem[\protect\citeauthoryear{Anderson, Bai, Bischof, Blackford, Dongarra,
  Du~Croz, Greenbaum, Hammarling, McKenney, and Sorensen}{Anderson
  et~al\mbox{.}}{1999}]%
        {anderson1999}
\bibfield{author}{\bibinfo{person}{Edward Anderson}, \bibinfo{person}{Zhaojun
  Bai}, \bibinfo{person}{Christian Bischof}, \bibinfo{person}{Susan Blackford},
  \bibinfo{person}{Jack Dongarra}, \bibinfo{person}{Jeremy Du~Croz},
  \bibinfo{person}{Anne Greenbaum}, \bibinfo{person}{Sven Hammarling},
  \bibinfo{person}{A. McKenney}, {and} \bibinfo{person}{D. Sorensen}.}
  \bibinfo{year}{1999}\natexlab{}.
\newblock \bibinfo{booktitle}{\emph{{LAPACK} {U}sers' guide}}.
  Vol.~\bibinfo{volume}{9}.
\newblock \bibinfo{publisher}{SIAM}.
\newblock


\bibitem[\protect\citeauthoryear{Baader and Nipkow}{Baader and Nipkow}{1999}]%
        {Baader:1999uu}
\bibfield{author}{\bibinfo{person}{Franz Baader} {and} \bibinfo{person}{Tobias
  Nipkow}.} \bibinfo{year}{1999}\natexlab{}.
\newblock \bibinfo{booktitle}{\emph{{Term Rewriting and All That}}}.
  Vol.~\bibinfo{volume}{31}.
\newblock \bibinfo{publisher}{Cambridge University Press}.
\newblock


\bibitem[\protect\citeauthoryear{Barthels, Copik, and Bientinesi}{Barthels
  et~al\mbox{.}}{2018}]%
        {barthels2018cgo}
\bibfield{author}{\bibinfo{person}{Henrik Barthels}, \bibinfo{person}{Marcin
  Copik}, {and} \bibinfo{person}{Paolo Bientinesi}.}
  \bibinfo{year}{2018}\natexlab{}.
\newblock \showarticletitle{{T}he {G}eneralized {M}atrix {C}hain {A}lgorithm}.
  In \bibinfo{booktitle}{\emph{Proceedings of 2018 IEEE/ACM International
  Symposium on Code Generation and Optimization}}. \bibinfo{address}{Vienna,
  Austria}, \bibinfo{pages}{138--148}.
\newblock
\urldef\tempurl%
\url{https://doi.org/10.1145/3168804}
\showDOI{\tempurl}


\bibitem[\protect\citeauthoryear{Barthels, Psarras, and Bientinesi}{Barthels
  et~al\mbox{.}}{2019}]%
        {barthels2019arxiv}
\bibfield{author}{\bibinfo{person}{Henrik Barthels}, \bibinfo{person}{Christos
  Psarras}, {and} \bibinfo{person}{Paolo Bientinesi}.}
  \bibinfo{year}{2019}\natexlab{}.
\newblock \showarticletitle{{A}utomatic {G}eneration of {E}fficient {L}inear
  {A}lgebra {P}rograms}.
\newblock \bibinfo{journal}{\emph{CoRR}}  \bibinfo{volume}{abs/1907.02778}
  (\bibinfo{year}{2019}).
\newblock
\showeprint[arXiv]{1907.02778}
\urldef\tempurl%
\url{http://arxiv.org/abs/1907.02778}
\showURL{%
\tempurl}
\newblock
\shownote{To appear in the Proceedings of the Platform for Advanced Scientific
  Computing Conference (PASC20).}


\bibitem[\protect\citeauthoryear{Baumgartner, Auer, Bernholdt, Bibireata,
  Choppella, Cociorva, Gao, Harrison, Hirata, Krishnamoorthy, Krishnan, Lam,
  Lu, Nooijen, Pitzer, Ramanujam, Sadayappan, and Sibiryakov}{Baumgartner
  et~al\mbox{.}}{2005}]%
        {Baumgartner:2005dq}
\bibfield{author}{\bibinfo{person}{Gerald Baumgartner},
  \bibinfo{person}{Alexander~A. Auer}, \bibinfo{person}{David~E. Bernholdt},
  \bibinfo{person}{Alina Bibireata}, \bibinfo{person}{Venkatesh Choppella},
  \bibinfo{person}{Daniel Cociorva}, \bibinfo{person}{Xiaoyang Gao},
  \bibinfo{person}{Robert~J. Harrison}, \bibinfo{person}{So Hirata},
  \bibinfo{person}{Sriram Krishnamoorthy}, \bibinfo{person}{Sandhya Krishnan},
  \bibinfo{person}{Chi-Chung Lam}, \bibinfo{person}{Qingda Lu},
  \bibinfo{person}{Marcel Nooijen}, \bibinfo{person}{Russell~M. Pitzer},
  \bibinfo{person}{J. Ramanujam}, \bibinfo{person}{P. Sadayappan}, {and}
  \bibinfo{person}{Alexander Sibiryakov}.} \bibinfo{year}{2005}\natexlab{}.
\newblock \showarticletitle{{Synthesis of High-Performance Parallel Programs
  for a Class of ab Initio Quantum Chemistry Models.}}
\newblock \bibinfo{journal}{\emph{Proc. IEEE}} \bibinfo{volume}{93},
  \bibinfo{number}{2} (\bibinfo{year}{2005}), \bibinfo{pages}{276--292}.
\newblock


\bibitem[\protect\citeauthoryear{Bezanson, Chen, Chung, Karpinski, Shah, Vitek,
  and Zoubritzky}{Bezanson et~al\mbox{.}}{2018}]%
        {Bezanson:2018ip}
\bibfield{author}{\bibinfo{person}{Jeff Bezanson}, \bibinfo{person}{Jiahao
  Chen}, \bibinfo{person}{Benjamin Chung}, \bibinfo{person}{Stefan Karpinski},
  \bibinfo{person}{Viral~B. Shah}, \bibinfo{person}{Jan Vitek}, {and}
  \bibinfo{person}{Lionel Zoubritzky}.} \bibinfo{year}{2018}\natexlab{}.
\newblock \showarticletitle{{Julia: Dynamism and Performance Reconciled by
  Design}}.
\newblock \bibinfo{journal}{\emph{Proceedings of the ACM on Programming
  Languages}} \bibinfo{volume}{2}, \bibinfo{number}{OOPSLA}
  (\bibinfo{date}{Oct.} \bibinfo{year}{2018}), \bibinfo{pages}{120--23}.
\newblock


\bibitem[\protect\citeauthoryear{Bientinesi, Gunter, and Geijn}{Bientinesi
  et~al\mbox{.}}{2008}]%
        {bientinesi2008}
\bibfield{author}{\bibinfo{person}{Paolo Bientinesi}, \bibinfo{person}{Brian
  Gunter}, {and} \bibinfo{person}{Robert A. van~de Geijn}.}
  \bibinfo{year}{2008}\natexlab{}.
\newblock \showarticletitle{{F}amilies of {A}lgorithms {R}elated to the
  {I}nversion of a {S}ymmetric {P}ositive {D}efinite {M}atrix}.
\newblock \bibinfo{journal}{\emph{ACM Trans. Math. Softw.}}
  \bibinfo{volume}{35}, \bibinfo{number}{1}, Article \bibinfo{articleno}{3}
  (\bibinfo{date}{July} \bibinfo{year}{2008}), \bibinfo{numpages}{22}~pages.
\newblock
\showISSN{0098-3500}
\urldef\tempurl%
\url{https://doi.org/10.1145/1377603.1377606}
\showDOI{\tempurl}


\bibitem[\protect\citeauthoryear{Bientinesi, Quintana-Ort{\'\i}, and van~de
  Geijn}{Bientinesi et~al\mbox{.}}{2005}]%
        {Bientinesi:2005hu}
\bibfield{author}{\bibinfo{person}{Paolo Bientinesi},
  \bibinfo{person}{Enrique~S. Quintana-Ort{\'\i}}, {and}
  \bibinfo{person}{Robert~A. van~de Geijn}.} \bibinfo{year}{2005}\natexlab{}.
\newblock \showarticletitle{{Representing linear algebra algorithms in code:
  the FLAME application program interfaces}}.
\newblock \bibinfo{journal}{\emph{ACM Trans. Math. Software}}
  \bibinfo{volume}{31}, \bibinfo{number}{1} (\bibinfo{date}{March}
  \bibinfo{year}{2005}), \bibinfo{pages}{27--59}.
\newblock


\bibitem[\protect\citeauthoryear{Bientinesi and van~de Geijn}{Bientinesi and
  van~de Geijn}{2011}]%
        {Bientinesi:2011bv}
\bibfield{author}{\bibinfo{person}{Paolo Bientinesi} {and}
  \bibinfo{person}{Robert~A. van~de Geijn}.} \bibinfo{year}{2011}\natexlab{}.
\newblock \showarticletitle{{Goal-Oriented and Modular Stability Analysis.}}
\newblock \bibinfo{journal}{\emph{SIAM J. Matrix Analysis Applications}}
  \bibinfo{volume}{32}, \bibinfo{number}{1} (\bibinfo{year}{2011}),
  \bibinfo{pages}{286--308}.
\newblock


\bibitem[\protect\citeauthoryear{Chung, Chung, Slagel, and Tenorio}{Chung
  et~al\mbox{.}}{2017}]%
        {Chung:2017ws}
\bibfield{author}{\bibinfo{person}{Julianne Chung}, \bibinfo{person}{Matthias
  Chung}, \bibinfo{person}{J.~Tanner Slagel}, {and} \bibinfo{person}{Luis
  Tenorio}.} \bibinfo{year}{2017}\natexlab{}.
\newblock \showarticletitle{{Stochastic Newton and Quasi-Newton Methods for
  Large Linear Least-squares Problems.}}
\newblock \bibinfo{journal}{\emph{CoRR}}  \bibinfo{volume}{math.NA}
  (\bibinfo{year}{2017}).
\newblock


\bibitem[\protect\citeauthoryear{Dick}{Dick}{1991}]%
        {Dick:1991jh}
\bibfield{author}{\bibinfo{person}{Alan J.~J. Dick}.}
  \bibinfo{year}{1991}\natexlab{}.
\newblock \showarticletitle{{An Introduction to Knuth-Bendix Completion}}.
\newblock \bibinfo{journal}{\emph{Comput. J.}} \bibinfo{volume}{34},
  \bibinfo{number}{1} (\bibinfo{date}{Jan.} \bibinfo{year}{1991}),
  \bibinfo{pages}{2--15}.
\newblock


\bibitem[\protect\citeauthoryear{Ding and Selesnick}{Ding and
  Selesnick}{2016}]%
        {ding2016}
\bibfield{author}{\bibinfo{person}{Yin Ding} {and} \bibinfo{person}{Ivan~W.
  Selesnick}.} \bibinfo{year}{2016}\natexlab{}.
\newblock \showarticletitle{{S}parsity-{B}ased {C}orrection of {E}xponential
  {A}rtifacts}.
\newblock \bibinfo{journal}{\emph{Signal Processing}}  \bibinfo{volume}{120}
  (\bibinfo{year}{2016}), \bibinfo{pages}{236--248}.
\newblock


\bibitem[\protect\citeauthoryear{Dongarra, Du~Croz, Hammarling, and
  Duff}{Dongarra et~al\mbox{.}}{1990}]%
        {dongarra1990}
\bibfield{author}{\bibinfo{person}{Jack~J. Dongarra}, \bibinfo{person}{Jeremy
  Du~Croz}, \bibinfo{person}{Sven Hammarling}, {and} \bibinfo{person}{Iain~S.
  Duff}.} \bibinfo{year}{1990}\natexlab{}.
\newblock \showarticletitle{{A} set of {L}evel 3 {B}asic {L}inear {A}lgebra
  {S}ubprograms}.
\newblock \bibinfo{journal}{\emph{ACM Transactions on Mathematical Software
  (TOMS)}} \bibinfo{volume}{16}, \bibinfo{number}{1} (\bibinfo{year}{1990}),
  \bibinfo{pages}{1--17}.
\newblock


\bibitem[\protect\citeauthoryear{Fabregat-Traver and
  Bientinesi}{Fabregat-Traver and Bientinesi}{2011a}]%
        {FabregatTraver:2011km}
\bibfield{author}{\bibinfo{person}{Diego Fabregat-Traver} {and}
  \bibinfo{person}{Paolo Bientinesi}.} \bibinfo{year}{2011}\natexlab{a}.
\newblock \showarticletitle{{Automatic Generation of Loop-Invariants for Matrix
  Operations.}}
\newblock \bibinfo{journal}{\emph{ICCSA Workshops}} (\bibinfo{year}{2011}),
  \bibinfo{pages}{82--92}.
\newblock


\bibitem[\protect\citeauthoryear{Fabregat-Traver and
  Bientinesi}{Fabregat-Traver and Bientinesi}{2011b}]%
        {FabregatTraver:2011gu}
\bibfield{author}{\bibinfo{person}{Diego Fabregat-Traver} {and}
  \bibinfo{person}{Paolo Bientinesi}.} \bibinfo{year}{2011}\natexlab{b}.
\newblock \showarticletitle{{Knowledge-Based Automatic Generation of
  Partitioned Matrix Expressions.}}
\newblock \bibinfo{journal}{\emph{CASC}} \bibinfo{volume}{6885},
  \bibinfo{number}{4} (\bibinfo{year}{2011}), \bibinfo{pages}{144--157}.
\newblock


\bibitem[\protect\citeauthoryear{Fabregat-Traver and
  Bientinesi}{Fabregat-Traver and Bientinesi}{2013}]%
        {fabregat-traver2013a}
\bibfield{author}{\bibinfo{person}{Diego Fabregat-Traver} {and}
  \bibinfo{person}{Paolo Bientinesi}.} \bibinfo{year}{2013}\natexlab{}.
\newblock \showarticletitle{{A} {D}omain-{S}pecific {C}ompiler for {L}inear
  {A}lgebra {O}perations}. In \bibinfo{booktitle}{\emph{High Performance
  Computing for Computational Science -- VECPAR 2010}}
  \emph{(\bibinfo{series}{Lecture Notes in Computer Science})},
  \bibfield{editor}{\bibinfo{person}{O.~Marques M.~Dayde} {and}
  \bibinfo{person}{K.~Nakajima}} (Eds.), Vol.~\bibinfo{volume}{7851}.
  \bibinfo{publisher}{Springer}, \bibinfo{address}{Heidelberg},
  \bibinfo{pages}{346--361}.
\newblock


\bibitem[\protect\citeauthoryear{Franchetti, Low, Popovici, Veras, Spampinato,
  Johnson, P{\"u}schel, Hoe, and Moura}{Franchetti et~al\mbox{.}}{2018}]%
        {Franchetti:eq}
\bibfield{author}{\bibinfo{person}{Franz Franchetti}, \bibinfo{person}{Tze~Meng
  Low}, \bibinfo{person}{Doru~Thom Popovici}, \bibinfo{person}{Richard~M.
  Veras}, \bibinfo{person}{Daniele~G. Spampinato}, \bibinfo{person}{Jeremy~R.
  Johnson}, \bibinfo{person}{Markus P{\"u}schel}, \bibinfo{person}{James~C.
  Hoe}, {and} \bibinfo{person}{Jose M.~F. Moura}.}
  \bibinfo{year}{2018}\natexlab{}.
\newblock \showarticletitle{{SPIRAL: Extreme Performance Portability}}.
\newblock \bibinfo{journal}{\emph{Proc. IEEE}} \bibinfo{volume}{106},
  \bibinfo{number}{11} (\bibinfo{date}{Nov.} \bibinfo{year}{2018}),
  \bibinfo{pages}{1935--1968}.
\newblock


\bibitem[\protect\citeauthoryear{Frigo and Johnson}{Frigo and Johnson}{2005}]%
        {Frigo:2005cp}
\bibfield{author}{\bibinfo{person}{Matteo Frigo} {and}
  \bibinfo{person}{Steven~G. Johnson}.} \bibinfo{year}{2005}\natexlab{}.
\newblock \showarticletitle{{The Design and Implementation of FFTW3}}.
\newblock \bibinfo{journal}{\emph{Proc. IEEE}} \bibinfo{volume}{93},
  \bibinfo{number}{2} (\bibinfo{year}{2005}), \bibinfo{pages}{216--231}.
\newblock


\bibitem[\protect\citeauthoryear{Godbole}{Godbole}{1973}]%
        {godbole1973}
\bibfield{author}{\bibinfo{person}{Sadashiva~S. Godbole}.}
  \bibinfo{year}{1973}\natexlab{}.
\newblock \showarticletitle{{O}n {E}fficient {C}omputation of {M}atrix {C}hain
  {P}roducts}.
\newblock \bibinfo{journal}{\emph{IEEE Trans. Comput.}} \bibinfo{volume}{C-22},
  \bibinfo{number}{9} (\bibinfo{date}{Sept} \bibinfo{year}{1973}),
  \bibinfo{pages}{864--866}.
\newblock
\showISSN{0018-9340}
\urldef\tempurl%
\url{https://doi.org/10.1109/TC.1973.5009182}
\showDOI{\tempurl}


\bibitem[\protect\citeauthoryear{Golub, Hansen, and O'Leary}{Golub
  et~al\mbox{.}}{2006}]%
        {Golub:2006hl}
\bibfield{author}{\bibinfo{person}{Gene~H. Golub},
  \bibinfo{person}{Per~Christian Hansen}, {and} \bibinfo{person}{Dianne~P.
  O'Leary}.} \bibinfo{year}{2006}\natexlab{}.
\newblock \showarticletitle{{Tikhonov Regularization and Total Least Squares}}.
\newblock \bibinfo{journal}{\emph{SIAM J. Matrix Anal. Appl.}}
  \bibinfo{volume}{21}, \bibinfo{number}{1} (\bibinfo{date}{July}
  \bibinfo{year}{2006}), \bibinfo{pages}{185--194}.
\newblock


\bibitem[\protect\citeauthoryear{Gomez and Scott}{Gomez and Scott}{1998}]%
        {gomez1998}
\bibfield{author}{\bibinfo{person}{Claude Gomez} {and} \bibinfo{person}{Tony
  Scott}.} \bibinfo{year}{1998}\natexlab{}.
\newblock \showarticletitle{{M}aple {P}rograms for {G}enerating {E}fficient
  {FORTRAN} {C}ode for {S}erial and {V}ectorised {M}achines}.
\newblock \bibinfo{journal}{\emph{Computer Physics Communications}}
  \bibinfo{volume}{115}, \bibinfo{number}{2} (\bibinfo{year}{1998}),
  \bibinfo{pages}{548--562}.
\newblock


\bibitem[\protect\citeauthoryear{Gower and Richt{\'a}rik}{Gower and
  Richt{\'a}rik}{2017}]%
        {Gower:2017bq}
\bibfield{author}{\bibinfo{person}{Robert~M. Gower} {and}
  \bibinfo{person}{Peter Richt{\'a}rik}.} \bibinfo{year}{2017}\natexlab{}.
\newblock \showarticletitle{{Randomized Quasi-Newton Updates Are Linearly
  Convergent Matrix Inversion Algorithms.}}
\newblock \bibinfo{journal}{\emph{SIAM J. Matrix Analysis Applications}}
  \bibinfo{volume}{38}, \bibinfo{number}{4} (\bibinfo{year}{2017}),
  \bibinfo{pages}{1380--1409}.
\newblock


\bibitem[\protect\citeauthoryear{Guennebaud, Jacob, et~al\mbox{.}}{Guennebaud
  et~al\mbox{.}}{2010}]%
        {eigenweb}
\bibfield{author}{\bibinfo{person}{Ga\"{e}l Guennebaud},
  \bibinfo{person}{Beno\^{i}t Jacob}, {et~al\mbox{.}}}
  \bibinfo{year}{2010}\natexlab{}.
\newblock \bibinfo{title}{{E}igen v3}.
\newblock \bibinfo{howpublished}{\url{http://eigen.tuxfamily.org}}.
\newblock


\bibitem[\protect\citeauthoryear{Gunnels, Gustavson, Henry, and van~de
  Geijn}{Gunnels et~al\mbox{.}}{2001}]%
        {Gunnels:2001gi}
\bibfield{author}{\bibinfo{person}{John~A. Gunnels}, \bibinfo{person}{Fred~G.
  Gustavson}, \bibinfo{person}{Greg~M. Henry}, {and} \bibinfo{person}{Robert~A.
  van~de Geijn}.} \bibinfo{year}{2001}\natexlab{}.
\newblock \showarticletitle{{FLAME: Formal Linear Algebra Methods
  Environment}}.
\newblock \bibinfo{journal}{\emph{ACM Trans. Math. Software}}
  \bibinfo{volume}{27}, \bibinfo{number}{4} (\bibinfo{date}{Dec.}
  \bibinfo{year}{2001}), \bibinfo{pages}{422--455}.
\newblock


\bibitem[\protect\citeauthoryear{Higham}{Higham}{1996}]%
        {higham1996}
\bibfield{author}{\bibinfo{person}{Nicholas~J. Higham}.}
  \bibinfo{year}{1996}\natexlab{}.
\newblock \bibinfo{booktitle}{\emph{{A}ccuracy and {S}tability of {N}umerical
  {A}lgorithms}}.
\newblock \bibinfo{publisher}{SIAM}, \bibinfo{address}{Philadelphia, PA, USA}.
\newblock


\bibitem[\protect\citeauthoryear{Higham}{Higham}{2008}]%
        {higham2008}
\bibfield{author}{\bibinfo{person}{Nicholas~J. Higham}.}
  \bibinfo{year}{2008}\natexlab{}.
\newblock \bibinfo{booktitle}{\emph{{F}unctions of {M}atrices: {T}heory and
  {C}omputation}}.
\newblock \bibinfo{publisher}{SIAM}, \bibinfo{address}{Philadelphia, PA, USA}.
\newblock
\showISBNx{978-0-898716-46-7}


\bibitem[\protect\citeauthoryear{Hu and Shing}{Hu and Shing}{1982}]%
        {hu1982}
\bibfield{author}{\bibinfo{person}{T.C. Hu} {and} \bibinfo{person}{M.T.
  Shing}.} \bibinfo{year}{1982}\natexlab{}.
\newblock \showarticletitle{{C}omputation of {M}atrix {C}hain {P}roducts.
  {P}art {I}}.
\newblock \bibinfo{journal}{\emph{SIAM J. Comput.}} \bibinfo{volume}{11},
  \bibinfo{number}{2} (\bibinfo{year}{1982}), \bibinfo{pages}{362--373}.
\newblock


\bibitem[\protect\citeauthoryear{Hu and Shing}{Hu and Shing}{1984}]%
        {hu1984}
\bibfield{author}{\bibinfo{person}{T.C. Hu} {and} \bibinfo{person}{M.T.
  Shing}.} \bibinfo{year}{1984}\natexlab{}.
\newblock \showarticletitle{{C}omputation of {M}atrix {C}hain {P}roducts.
  {P}art {II}}.
\newblock \bibinfo{journal}{\emph{SIAM J. Comput.}} \bibinfo{volume}{13},
  \bibinfo{number}{2} (\bibinfo{year}{1984}), \bibinfo{pages}{228--251}.
\newblock


\bibitem[\protect\citeauthoryear{Iakymchuk and Bientinesi}{Iakymchuk and
  Bientinesi}{2012}]%
        {iakymchuk2012}
\bibfield{author}{\bibinfo{person}{Roman Iakymchuk} {and}
  \bibinfo{person}{Paolo Bientinesi}.} \bibinfo{year}{2012}\natexlab{}.
\newblock \showarticletitle{{M}odeling {P}erformance through
  {M}emory-{S}talls}.
\newblock \bibinfo{journal}{\emph{ACM SIGMETRICS Performance Evaluation
  Review}} \bibinfo{volume}{40}, \bibinfo{number}{2} (\bibinfo{date}{Oct.}
  \bibinfo{year}{2012}), \bibinfo{pages}{86--91}.
\newblock


\bibitem[\protect\citeauthoryear{Iglberger, Hager, Treibig, and
  R{\"u}de}{Iglberger et~al\mbox{.}}{2012}]%
        {Iglberger:2012hb}
\bibfield{author}{\bibinfo{person}{Klaus Iglberger}, \bibinfo{person}{Georg
  Hager}, \bibinfo{person}{Jan Treibig}, {and} \bibinfo{person}{Ulrich
  R{\"u}de}.} \bibinfo{year}{2012}\natexlab{}.
\newblock \showarticletitle{{Expression Templates Revisited: A Performance
  Analysis of Current Methodologies.}}
\newblock \bibinfo{journal}{\emph{SIAM J. Scientific Computing}}
  \bibinfo{volume}{34}, \bibinfo{number}{2} (\bibinfo{year}{2012}),
  \bibinfo{pages}{C42--C69}.
\newblock


\bibitem[\protect\citeauthoryear{{Intel Corporation}}{{Intel
  Corporation}}{2019}]%
        {mkldoc}
\bibfield{author}{\bibinfo{person}{{Intel Corporation}}.}
  \bibinfo{year}{2019}\natexlab{}.
\newblock \bibinfo{title}{{I}ntel\textregistered {M}ath {K}ernel {L}ibrary
  documentation}.
\newblock
  \bibinfo{howpublished}{\url{https://software.intel.com/en-us/mkl-reference-manual-for-c}}.
\newblock


\bibitem[\protect\citeauthoryear{Jim{\'e}nez and Marzal}{Jim{\'e}nez and
  Marzal}{1999}]%
        {Jimenez:1999cd}
\bibfield{author}{\bibinfo{person}{V{\'\i}ctor~M. Jim{\'e}nez} {and}
  \bibinfo{person}{Andr{\'e}s Marzal}.} \bibinfo{year}{1999}\natexlab{}.
\newblock \showarticletitle{{Computing the K Shortest Paths - A New Algorithm
  and an Experimental Comparison.}}
\newblock \bibinfo{journal}{\emph{Algorithm Engineering}}
  \bibinfo{volume}{1668}, \bibinfo{number}{Chapter 4} (\bibinfo{year}{1999}),
  \bibinfo{pages}{15--29}.
\newblock


\bibitem[\protect\citeauthoryear{Kabal}{Kabal}{2011}]%
        {Kabal:2011wr}
\bibfield{author}{\bibinfo{person}{Peter Kabal}.}
  \bibinfo{year}{2011}\natexlab{}.
\newblock \showarticletitle{{Minimum Mean-Square Error Filtering:
  Autocorrelation/Covariance, General Delays, and Multirate Systems}}.
\newblock  (\bibinfo{year}{2011}).
\newblock


\bibitem[\protect\citeauthoryear{Kalman}{Kalman}{1960}]%
        {Kalman:1960ii}
\bibfield{author}{\bibinfo{person}{Rudolf~Emil Kalman}.}
  \bibinfo{year}{1960}\natexlab{}.
\newblock \showarticletitle{{A New Approach to Linear Filtering and Prediction
  Problems}}.
\newblock \bibinfo{journal}{\emph{Journal of basic Engineering}}
  \bibinfo{volume}{82}, \bibinfo{number}{1} (\bibinfo{date}{March}
  \bibinfo{year}{1960}), \bibinfo{pages}{35--45}.
\newblock


\bibitem[\protect\citeauthoryear{Krebber}{Krebber}{2017}]%
        {Krebber:2017tp}
\bibfield{author}{\bibinfo{person}{Manuel Krebber}.}
  \bibinfo{year}{2017}\natexlab{}.
\newblock \showarticletitle{{Non-linear Associative-Commutative Many-to-One
  Pattern Matching with Sequence Variables.}}
\newblock \bibinfo{journal}{\emph{CoRR}}  \bibinfo{volume}{cs.SC}
  (\bibinfo{year}{2017}).
\newblock


\bibitem[\protect\citeauthoryear{Krebber and Barthels}{Krebber and
  Barthels}{2018}]%
        {krebber2018joss}
\bibfield{author}{\bibinfo{person}{Manuel Krebber} {and}
  \bibinfo{person}{Henrik Barthels}.} \bibinfo{year}{2018}\natexlab{}.
\newblock \showarticletitle{{M}atch{P}y: {P}attern {M}atching in {P}ython}.
\newblock \bibinfo{journal}{\emph{Journal of Open Source Software}}
  \bibinfo{volume}{3}, \bibinfo{number}{26} (\bibinfo{date}{June}
  \bibinfo{year}{2018}), \bibinfo{pages}{2}.
\newblock
\urldef\tempurl%
\url{https://doi.org/10.21105/joss.00670}
\showDOI{\tempurl}


\bibitem[\protect\citeauthoryear{Krebber, Barthels, and Bientinesi}{Krebber
  et~al\mbox{.}}{2017a}]%
        {krebber2017pyhpc}
\bibfield{author}{\bibinfo{person}{Manuel Krebber}, \bibinfo{person}{Henrik
  Barthels}, {and} \bibinfo{person}{Paolo Bientinesi}.}
  \bibinfo{year}{2017}\natexlab{a}.
\newblock \showarticletitle{{E}fficient {P}attern {M}atching in {P}ython}. In
  \bibinfo{booktitle}{\emph{Proceedings of the 7th Workshop on Python for
  High-Performance and Scientific Computing}}. \bibinfo{address}{Denver,
  Colorado}.
\newblock
\urldef\tempurl%
\url{https://doi.org/10.1145/3149869.3149871}
\showDOI{\tempurl}
\showeprint[arXiv]{1710.00077}
\newblock
\shownote{In conjunction with SC17: The International Conference for High
  Performance Computing, Networking, Storage and Analysis.}


\bibitem[\protect\citeauthoryear{Krebber, Barthels, and Bientinesi}{Krebber
  et~al\mbox{.}}{2017b}]%
        {krebber2017scipy}
\bibfield{author}{\bibinfo{person}{Manuel Krebber}, \bibinfo{person}{Henrik
  Barthels}, {and} \bibinfo{person}{Paolo Bientinesi}.}
  \bibinfo{year}{2017}\natexlab{b}.
\newblock \showarticletitle{{M}atch{P}y: {A} {P}attern {M}atching {L}ibrary}.
  In \bibinfo{booktitle}{\emph{Proceedings of the 15th Python in Science
  Conference}}. \bibinfo{address}{Austin, Texas}.
\newblock
\showeprint[arXiv]{1710.06915}
\urldef\tempurl%
\url{https://arxiv.org/abs/1710.06915}
\showURL{%
\tempurl}


\bibitem[\protect\citeauthoryear{Marker, Poulson, Batory, and van~de
  Geijn}{Marker et~al\mbox{.}}{2012}]%
        {marker2012}
\bibfield{author}{\bibinfo{person}{Bryan Marker}, \bibinfo{person}{Jack
  Poulson}, \bibinfo{person}{Don Batory}, {and} \bibinfo{person}{Robert van~de
  Geijn}.} \bibinfo{year}{2012}\natexlab{}.
\newblock \showarticletitle{{D}esigning {L}inear {A}lgebra {A}lgorithms by
  {T}ransformation: {M}echanizing the {E}xpert {D}eveloper}.
\newblock In \bibinfo{booktitle}{\emph{High Performance Computing for
  Computational Science-VECPAR 2012}}. \bibinfo{publisher}{Springer},
  \bibinfo{pages}{362--378}.
\newblock


\bibitem[\protect\citeauthoryear{Marker, Schatz, Matthews, Dillig, van~de
  Geijn, and Batory}{Marker et~al\mbox{.}}{2015}]%
        {marker2015}
\bibfield{author}{\bibinfo{person}{Bryan Marker}, \bibinfo{person}{Martin
  Schatz}, \bibinfo{person}{Devin Matthews}, \bibinfo{person}{Isil Dillig},
  \bibinfo{person}{Robert van~de Geijn}, {and} \bibinfo{person}{Don Batory}.}
  \bibinfo{year}{2015}\natexlab{}.
\newblock \bibinfo{booktitle}{\emph{{D}xter: {A}n {E}xtensible {T}ool for
  {O}ptimal {D}ataflow {P}rogram {G}eneration}}.
\newblock \bibinfo{type}{{T}echnical {R}eport}. \bibinfo{institution}{Technical
  report, Technical Report TR-15-03, The University of Texas at Austin}.
\newblock


\bibitem[\protect\citeauthoryear{Muchnick}{Muchnick}{1997}]%
        {Muchnick:1997wv}
\bibfield{author}{\bibinfo{person}{Steven~S. Muchnick}.}
  \bibinfo{year}{1997}\natexlab{}.
\newblock \bibinfo{booktitle}{\emph{{Advanced Compiler Design and
  Implementation}}}.
\newblock \bibinfo{publisher}{Morgan Kaufmann}.
\newblock


\bibitem[\protect\citeauthoryear{Ni{\~{n}}o, Sandu, and Deng}{Ni{\~{n}}o
  et~al\mbox{.}}{2016}]%
        {nino2016}
\bibfield{author}{\bibinfo{person}{Elias~D. Ni{\~{n}}o},
  \bibinfo{person}{Adrian Sandu}, {and} \bibinfo{person}{Xinwei Deng}.}
  \bibinfo{year}{2016}\natexlab{}.
\newblock \showarticletitle{{A} {P}arallel {I}mplementation of the {E}nsemble
  {K}alman {F}ilter {B}ased on {M}odified {C}holesky {D}ecomposition}.
\newblock \bibinfo{journal}{\emph{CoRR}}  \bibinfo{volume}{abs/1606.00807}
  (\bibinfo{year}{2016}).
\newblock
\urldef\tempurl%
\url{http://arxiv.org/abs/1606.00807}
\showURL{%
\tempurl}


\bibitem[\protect\citeauthoryear{Peise and Bientinesi}{Peise and
  Bientinesi}{2012}]%
        {peise2012}
\bibfield{author}{\bibinfo{person}{Elmar Peise} {and} \bibinfo{person}{Paolo
  Bientinesi}.} \bibinfo{year}{2012}\natexlab{}.
\newblock \showarticletitle{{P}erformance {M}odeling for {D}ense {L}inear
  {A}lgebra}. In \bibinfo{booktitle}{\emph{Proceedings of the 2012 SC
  Companion: High Performance Computing, Networking Storage and Analysis
  (PMBS12)}} \emph{(\bibinfo{series}{SCC '12})}. \bibinfo{publisher}{IEEE
  Computer Society}, \bibinfo{address}{Washington, DC, USA},
  \bibinfo{pages}{406--416}.
\newblock


\bibitem[\protect\citeauthoryear{Peise and Bientinesi}{Peise and
  Bientinesi}{2014}]%
        {Peise:2014fr}
\bibfield{author}{\bibinfo{person}{Elmar Peise} {and} \bibinfo{person}{Paolo
  Bientinesi}.} \bibinfo{year}{2014}\natexlab{}.
\newblock \showarticletitle{{A Study on the Influence of Caching: Sequences of
  Dense Linear Algebra Kernels}}. In \bibinfo{booktitle}{\emph{High Performance
  Computing for Computational Science -- VECPAR 2014}}.
  \bibinfo{publisher}{Springer, Cham}, \bibinfo{pages}{245--258}.
\newblock


\bibitem[\protect\citeauthoryear{Pelegri-Llopart and Graham}{Pelegri-Llopart
  and Graham}{1988}]%
        {pelegri1988}
\bibfield{author}{\bibinfo{person}{Eduardo Pelegri-Llopart} {and}
  \bibinfo{person}{Susan~L Graham}.} \bibinfo{year}{1988}\natexlab{}.
\newblock \showarticletitle{{O}ptimal {C}ode {G}eneration for {E}xpression
  {T}rees: {A}n {A}pplication {BURS} {T}heory}. In
  \bibinfo{booktitle}{\emph{Proceedings of the 15th ACM SIGPLAN-SIGACT
  symposium on Principles of programming languages}}. ACM,
  \bibinfo{pages}{294--308}.
\newblock


\bibitem[\protect\citeauthoryear{Psarras, Barthels, and Bientinesi}{Psarras
  et~al\mbox{.}}{2019}]%
        {psarras2019arxiv}
\bibfield{author}{\bibinfo{person}{Christos Psarras}, \bibinfo{person}{Henrik
  Barthels}, {and} \bibinfo{person}{Paolo Bientinesi}.}
  \bibinfo{year}{2019}\natexlab{}.
\newblock \showarticletitle{{The Linear Algebra Mapping Problem}}.
\newblock \bibinfo{journal}{\emph{CoRR}}  \bibinfo{volume}{abs/1911.09421}
  (\bibinfo{year}{2019}).
\newblock
\showeprint[arxiv]{cs.MS/1911.09421}
\urldef\tempurl%
\url{http://arxiv.org/abs/1911.09421}
\showURL{%
\tempurl}


\bibitem[\protect\citeauthoryear{Sanderson}{Sanderson}{2010}]%
        {sanderson2010}
\bibfield{author}{\bibinfo{person}{Conrad Sanderson}.}
  \bibinfo{year}{2010}\natexlab{}.
\newblock \showarticletitle{{A}rmadillo: {A}n {O}pen {S}ource {C}++ {L}inear
  {A}lgebra {L}ibrary for {F}ast {P}rototyping and {C}omputationally
  {I}ntensive {E}xperiments}.
\newblock  (\bibinfo{year}{2010}).
\newblock


\bibitem[\protect\citeauthoryear{Sethi and Ullman}{Sethi and Ullman}{1970}]%
        {sethi1970}
\bibfield{author}{\bibinfo{person}{Ravi Sethi} {and}
  \bibinfo{person}{Jeffrey~D. Ullman}.} \bibinfo{year}{1970}\natexlab{}.
\newblock \showarticletitle{{T}he {G}eneration of {O}ptimal {C}ode for
  {A}rithmetic {E}xpressions}.
\newblock \bibinfo{journal}{\emph{Journal of the ACM (JACM)}}
  \bibinfo{volume}{17}, \bibinfo{number}{4} (\bibinfo{year}{1970}),
  \bibinfo{pages}{715--728}.
\newblock


\bibitem[\protect\citeauthoryear{Siek, Karlin, and Jessup}{Siek
  et~al\mbox{.}}{2008}]%
        {Siek:2008ij}
\bibfield{author}{\bibinfo{person}{Jeremy~G. Siek}, \bibinfo{person}{Ian
  Karlin}, {and} \bibinfo{person}{Elizabeth~R. Jessup}.}
  \bibinfo{year}{2008}\natexlab{}.
\newblock \showarticletitle{{Build to Order Linear Algebra Kernels}}. In
  \bibinfo{booktitle}{\emph{Distributed Processing Symposium (IPDPS)}}.
  \bibinfo{publisher}{IEEE}, \bibinfo{pages}{1--8}.
\newblock


\bibitem[\protect\citeauthoryear{Spampinato, Fabregat-Traver, Bientinesi, and
  P{\"u}schel}{Spampinato et~al\mbox{.}}{2018}]%
        {Spampinato:2018tz}
\bibfield{author}{\bibinfo{person}{Daniele~G. Spampinato},
  \bibinfo{person}{Diego Fabregat-Traver}, \bibinfo{person}{Paolo Bientinesi},
  {and} \bibinfo{person}{Markus P{\"u}schel}.} \bibinfo{year}{2018}\natexlab{}.
\newblock \showarticletitle{{Program Generation for Small-Scale Linear Algebra
  Applications.}}. In \bibinfo{booktitle}{\emph{International Symposium on Code
  Generation and Optimization (CGO)}}. \bibinfo{publisher}{ACM Press},
  \bibinfo{address}{Vienna, Austria}, \bibinfo{pages}{327--339}.
\newblock


\bibitem[\protect\citeauthoryear{Spampinato and P{\"u}schel}{Spampinato and
  P{\"u}schel}{2016}]%
        {spampinato2016}
\bibfield{author}{\bibinfo{person}{Daniele~G. Spampinato} {and}
  \bibinfo{person}{Markus P{\"u}schel}.} \bibinfo{year}{2016}\natexlab{}.
\newblock \showarticletitle{{A} {B}asic {L}inear {A}lgebra {C}ompiler for
  {S}tructured {M}atrices}. In \bibinfo{booktitle}{\emph{International
  Symposium on Code Generation and Optimization (CGO)}}.
  \bibinfo{pages}{117--127}.
\newblock


\bibitem[\protect\citeauthoryear{Straszak and Vishnoi}{Straszak and
  Vishnoi}{2015}]%
        {straszak2015}
\bibfield{author}{\bibinfo{person}{Damian Straszak} {and}
  \bibinfo{person}{Nisheeth~K. Vishnoi}.} \bibinfo{year}{2015}\natexlab{}.
\newblock \showarticletitle{{O}n a {N}atural {D}ynamics for {L}inear
  {P}rogramming}.
\newblock  (\bibinfo{year}{2015}).
\newblock
\showeprint[arXiv]{1511.07020}


\bibitem[\protect\citeauthoryear{Tate, Stepp, Tatlock, and Lerner}{Tate
  et~al\mbox{.}}{2009}]%
        {Tate:2009kz}
\bibfield{author}{\bibinfo{person}{Ross Tate}, \bibinfo{person}{Michael Stepp},
  \bibinfo{person}{Zachary Tatlock}, {and} \bibinfo{person}{Sorin Lerner}.}
  \bibinfo{year}{2009}\natexlab{}.
\newblock \showarticletitle{{Equality Saturation - A New Approach to
  Optimization}}. In \bibinfo{booktitle}{\emph{Proceedings of the 36th Annual
  ACM SIGPLAN-SIGACT Symposium on Principles of Programming Languages}}.
  \bibinfo{publisher}{ACM Press}, \bibinfo{address}{New York, NY, USA},
  \bibinfo{pages}{264--276}.
\newblock


\bibitem[\protect\citeauthoryear{{The MathWorks, Inc.}}{{The MathWorks,
  Inc.}}{2019}]%
        {matlabdoc}
\bibfield{author}{\bibinfo{person}{{The MathWorks, Inc.}}}
  \bibinfo{year}{2019}\natexlab{}.
\newblock \bibinfo{title}{{M}atlab documentation}.
\newblock \bibinfo{howpublished}{\url{http://www.mathworks.com/help/matlab}}.
\newblock


\bibitem[\protect\citeauthoryear{Tirer and Giryes}{Tirer and Giryes}{2017}]%
        {Tirer:2017uv}
\bibfield{author}{\bibinfo{person}{Tom Tirer} {and} \bibinfo{person}{Raja
  Giryes}.} \bibinfo{year}{2017}\natexlab{}.
\newblock \showarticletitle{{Image Restoration by Iterative Denoising and
  Backward Projections}}.
\newblock \bibinfo{journal}{\emph{arXiv.org}} (\bibinfo{date}{Oct.}
  \bibinfo{year}{2017}), \bibinfo{pages}{138--142}.
\newblock
\showeprint[arxiv]{cs.CV/1710.06647v1}


\end{thebibliography}

\appendix

\section{Example Problems}
\label{sec:exampleproblems}

A selection of the 25 application problems used in the experiments. Matrix properties: diagonal (DI), lower/upper triangular (LT/UT), symmetric positive definite (SPD), symmetric positive semi-definite (SPSD), symmetric (SYM).

\newcommand{\LT}{\text{LT}}
\newcommand{\UT}{\text{UT}}
\newcommand{\DI}{\text{DI}}
\newcommand{\SYM}{\text{SYM}}
\newcommand{\SPD}{\text{SPD}}
\newcommand{\SPSD}{\text{SPSD}}
\newcommand{\dims}[3]{$#1 \in \mathbb{R}^{#2 \times #3}$}
\newcommand{\idx}[1]{\parbox{0.4cm}{#1}}

\newcommand{\exprob}[3]{
\subsection{#1}#2#3
}

\exprob{Generalized Least Squares}{
\[b := (X^T M^{-1} X)^{-1} X^T M^{-1} y\]
}{
\dims{M}{n}{n}, \SPD{};
\dims{X}{n}{m};
\dims{y}{n}{1};\\
$n > m$;
$n = 2500$;
$m = 500$
}

\exprob{Optimization \cite{straszak2015}}{
\begin{align*}
  x_f &:= W A^T (AWA^T)^{-1} (b - Ax)\\
  x_o &:= W (A^T (AWA^T)^{-1} Ax - c)
\end{align*}
}{
\dims{A}{m}{n};
\dims{W}{n}{n}, \DI, \SPD;
\dims{b}{m}{1};
\dims{c}{n}{1};\\
$n > m$;
$n = 2000$;
$m = 1000$
}

\exprob{Signal Processing \cite{ding2016}}{
\[x := ( A^{-T} B^T B A^{-1} + R^T L R )^{-1} A^{-T} B^T B A^{-1} y\]
}{
\dims{A}{n}{n};
\dims{B}{n}{n};
\dims{R}{n-1}{n}, \UT;
\dims{L}{n-1}{n-1}, \DI;
\dims{y}{n}{1};\\
$n = 2000$
}

\exprob{Triangular Matrix Inversion \cite{bientinesi2008}}{
\begin{align*}
  X_{10} &:= L_{10} L_{00}^{-1} \\
  X_{20} &:= L_{20} + L_{22}^{-1} L_{21} L_{11}^{-1} L_{10} \\
  X_{11} &:= L_{11}^{-1} \\
  X_{21} &:= - L_{22}^{-1} L_{21}
\end{align*}
}{
\dims{L_{00}}{n}{n}, \LT;
\dims{L_{11}}{m}{m}, \LT;
\dims{L_{22}}{k}{k}, \LT;
\dims{L_{10}}{m}{n};
\dims{L_{20}}{k}{n};
\dims{L_{21}}{k}{m};\\
$n = 2000$;
$m = 200$;
$k = 2000$
}
\label{exprob:triinv}

\exprob{Ensemble Kalman Filter \cite{nino2016}}{
\[X^a := X^b + ( B^{-1} + H^T R^{-1} H )^{-1} ( Y - H X^b )\]
}{
\dims{B}{N}{N} \SPSD;
\dims{H}{m}{N};
\dims{R}{m}{m} \SPSD;
\dims{Y}{m}{N};
\dims{X^b}{n}{N};\\
$N = 200$;
$n = 2000$;
$m = 2000$
}

\exprob{Image Restoration \cite{Tirer:2017uv}}{
\[x_k := (H^T H + \lambda \sigma^{2} I_{n} )^{-1} ( H^T y +\lambda \sigma^{2}(v_{k-1} - u_{k-1}) )\]
}{
\dims{H}{m}{n};
\dims{y}{m}{1};
\dims{v_{k-1}}{n}{1};
\dims{u_{k-1}}{n}{1};
$\lambda > 0$;
$\sigma > 0$;\\
$n > m$;
$n = 5000$;
$m = 1000$
}

\exprob{Randomized Matrix Inversion \cite{Gower:2017bq}}{
\begin{align*}
  \Lambda &:= S (S^T A^T W A S)^{-1} S^T \\
  X_{k+1} &:= X_k + (I_n - X_k A^T) \Lambda A^T W
\end{align*}
}{
\dims{W}{n}{n}, \SPD;
\dims{S}{n}{q};
\dims{A}{n}{n};
\dims{X_k}{n}{n};\\
$q \ll n$;
$n = 5000$;
$q = 500$
}
\label{exprob:randinv}

\exprob{Randomized Matrix Inversion \cite{Gower:2017bq}}{
\[X_{k+1} := S(S^T A S)^{-1} S^T + (I_n - S(S^T A S)^{-1} S^T A) X_k (I_n - A S(S^T A S)^{-1} S^T)\]
}{
\dims{A}{n}{n}, \SPD;
\dims{W}{n}{n}, \SPD;
\dims{S}{n}{q};
\dims{X_k}{n}{n};\\
$q \ll n$;
$n = 5000$;
$q = 500$
}

\exprob{Stochastic Newton \cite{Chung:2017ws}}{
\[B_k := \frac{k}{k-1}B_{k-1} (I_n - A^T W_k ((k-1)I_l + W_k^T A B_{k-1} A^T W_k)^{-1} W_k^T A B_{k-1} )\]
}{
\dims{W_k}{m}{l};
\dims{A}{m}{n};
\dims{B_k}{n}{n}, \SPD;\\
$l < n \ll m$;
$l = 625$;
$n = 1000$;
$m = 5000$
}

\exprob{Tikhonov regularization \cite{Golub:2006hl}}{
\[x := (A^T A + \Gamma^T \Gamma)^{-1} A^T b\]
}{
\dims{A}{n}{m};
\dims{\Gamma}{m}{m};
\dims{b}{n}{1};\\
$n = 3000$;
$m = 200$
}
\label{exprob:tikhonov}

\exprob{Generalized Tikhonov regularization}{
\[x := (A^T P A + Q)^{-1} (A^T P b + Q x_0 )\]
}{
\dims{P}{n}{n}, \SPSD;
\dims{Q}{m}{m}, \SPSD;
\dims{x_0}{m}{1};
\dims{A}{n}{m};
\dims{\Gamma}{m}{m};
\dims{b}{n}{1};\\
$n = 3000$;
$m = 200$
}

\exprob{LMMSE estimator \cite{Kabal:2011wr}}{
\[x_\text{out} := C_X A^T (A C_X A^T + C_Z)^{-1} (y - A x) + x\]
}{
\dims{A}{m}{n};
\dims{C_X}{n}{n}, \SPSD;
\dims{C_Z}{m}{m}, \SPSD;
\dims{x}{n}{1};
\dims{y}{m}{1};\\
$n = 2000$;
$m = 1500$
}

\exprob{Kalman Filter \cite{Kalman:1960ii}}
{
\begin{align*}
	K_k   &:= P_{k-1} H^{T} ( H_{k} P_{k-1} H_k^T + R_k )^{-1} \\
	P_k   &:= \left( I - K_k H_k \right) P_{k-1} \\
	x_{k} &:= x_{k-1} + K_k ( z_k - H_{k} x_{k-1} )
\end{align*}
}{
\dims{K_k}{n}{m};
\dims{P_k}{n}{n}, \SPD;
\dims{H_k}{m}{n}, \SPD;
\dims{R_k}{m}{m}, \SPSD;
\dims{x_k}{n}{1};
\dims{z_k}{m}{1};\\
$n = 400$;
$m = 500$
}

\end{document}